%% file: paper.tex
\documentclass[useAMS,usenatbib]{mn2e}

\input{abbrev}
\input{mydefs}
\input{psfig.sty}
\usepackage{graphicx}
\usepackage{epsfig}
\usepackage{amssymb}
\usepackage{amsmath}
\usepackage{amsfonts}
\usepackage{txfonts}
\usepackage{multirow}
\usepackage{xcolor}
\usepackage{ulem}

\title[Weak Lensing: Finite Support Effects ]{Simulations of Weak Gravitational Lensing - II : Including Finite Support Effects in Cosmic Shear Covariance Matrices}
\author[Joachim Harnois-D\'{e}raps et al.]{Joachim Harnois-D\'{e}raps$^{1,2}$, Ludovic van Waerbeke$^{1}$
\thanks{E-mail: jharno@cita.utoronto.ca}\\
$^{1}$Department of Physics and Astronomy, University of British Columbia, V6T 1Z1, B.C., Canada\\
$^{2}$Canadian Institute for Theoretical Astrophysics, University of
Toronto, M5S 3H8, On., Canada\\}

\begin{document}

\date{\today}

\pagerange{\pageref{firstpage}--\pageref{lastpage}} \pubyear{2013}

\maketitle

\label{firstpage}

\begin{abstract}

Numerical N-body simulations play a central role in the assessment of weak gravitational lensing statistics, residual systematics and error analysis. 
In this paper, we investigate and quantify the impact of  finite simulation volume on weak lensing two- and four-point statistics. 
These {\it finite support} (FS) effects are modelled for several estimators,  simulation box sizes and source redshifts, and validated against a new large suite of 500 $N$-body
simulations.  The comparison reveals that our theoretical model is accurate to better than  5 per cent for the shear correlation function $\xi_{+}(\theta)$ and its error.
We find that the most important quantities for FS modelling is the ratio between the  measured angle $\theta$ and the 
angular size of the simulation box at the source redshift, $\theta_{box}(z_s)$, or the multipole equivalent $\ell / \ell_{box}(z_s)$. 
When this ratio reaches 0.1, independently of the source redshift,
the shear correlation function $\xi_+$ is suppressed by 5, 10, 20 and 25 percent for $L_{box}= 1000$,  $500$, $250$ and $147 \mathMpch$ respectively.
When it reaches 0.2, the suppression exceeds 25 percent even for the largest box. 
The same effect is observed in $\xi_{-}(\theta)$, but at much larger angles. 
This has important consequences for cosmological analyses using $N$-body simulations to calibrate the impact of non-linear gravitational clustering or
to estimate errors and systematics effects, and should not be overlooked.
We propose  simple semi-analytic solutions to correct for these finite box effects with and without the presence of survey masks, 
and the method can be generalized to any weak lensing estimator.
With the corrections applied, both weak lensing signals and errors can be made accurate at any angles even in simulations as small as $L_{box}= 147\mathMpch$. 
This offers a graceful solution to the important problem of estimating accurate covariance matrices for weak lensing studies: 
there is no need to run extra large simulation volumes, as long as the box effects are corrected.
We can instead concentrate our efforts  on modelling the small scales accurately, eventually with hydrodynamical simulations. 
From the same simulation suite, we revisit the existing non-Gaussian covariance matrix calibration of the shear correlation function, and propose a new one based on the {\small WMAP}9 + BAO + SN cosmology. Our calibration matrix is accurate at $20$ percent down to the arc minute scale, for source redshifts $0 < z <3$,
even for the far off-diagonal elements. 
We propose, for the first time, a parameterization for the full $\xi_-$ covariance matrix, also better than 20 percent for most elements.
\end{abstract}



\section{Introduction}
\label{sec:intro}

Weak gravitational lensing  has emerged as one of the key methods to 
constrain astrophysical and cosmological parameters. The technique studies the distortions in the images of background luminous sources by foreground mass,
and is therefore sensitive to the total matter content along and surrounding the photon's trajectories.
This allows us to measure and map the combined contributions of all matter (dark matter, baryons and neutrinos) in an unbiased way \citep[see][for a review]{2008PhR...462...67M}.

Recent results from the CFHT Lensing Survey \citep{2013MNRAS.433.2545E, 2012MNRAS.427..146H}
have shown the potential of weak lensing as a powerful cosmological and astrophysical probe in a fully controlled residual systematics environment. A non exhaustive list of key results include accurate measurements of galaxy luminosity and stellar mass functions \citep{2014MNRAS.437.2111V},
studies of the galaxy/dark matter environmental connection  \citep{2013MNRAS.431.1439G}, 
tests for the laws of gravity \citep{2013MNRAS.429.2249S}, large scale structure mass maps \citep{2013MNRAS.433.3373V} and it has placed competitive constraints on many cosmological parameters which includes the first tomographic analysis taking into account intrinsic alignment \citep{2013MNRAS.430.2200K, 2013MNRAS.432.2433H}.
These results show an impressive list of scientific results for a survey that is `only' 150 square degrees; the sky coverage from the upcoming analysis with the RCS2\footnote{{\tt www.rcslens.org}}, DES\footnote{{\tt www.darkenergysurvey.org}}, KiDS\footnote{{\tt kids.strw.leidenuniv.nl/}} and HSC\footnote{{\tt www.naoj.org/Projects/HSC/}}  
surveys is more than an order of magnitude larger, therefore significantly increasing the statistical precision.

 Mock galaxy catalogues, based on numerical simulations, are playing a central role in weak lensing studies. These simulations are needed for the testing and calibration of statistical estimators. This is particularly important in the non-linear gravitational clustering regime, where theoretical predictions cannot be done analytically with high precision. Equally important is the necessity to understand any "contamination" signals (e.g. intrinsic alignment of galaxies, source-lens correlations) in a realistic completely non-linear clustering environment. The accurate estimation of the sampling variance at small scale must also be performed with numerical simulations. As shown in \citet{2012MNRAS.427..146H}, this is an essential element in the quantification of residual systematics since 
{with the telescope's point spread function} cannot be neglected.
 All these aspects of the weak lensing analyses are essential for a reliable interpretation of the data and an accurate 
 treatment of the errors. $N$-body simulations is the best approach to achieve these multiple goals.

Simulations are always performed inside a finite cosmological volume, therefore, density fluctuations larger than the computation box are ignored. 
How much this affects the weak lensing measurements on and other cosmological observations was subject of many studies \citep{1994A&A...281..301C, 2006MNRAS.370..691P, 2009ApJ...701..945S}, but is not yet known with full precision.
In the context of precision cosmology, it is critical to quantify this effect {, including all its subtle ramifications, } for all the statistical estimators being used.
In this work, we investigate a novel  {consequence}  of the missing large scale modes, which 
we refer to as the {\it finite support} (FS) effect. We develop a general strategy to {include} all known finite box effects in the covariance matrices based on pure dark matter $N$-body simulations,
and we provide a simple recipe for its implementation on cosmic shear data analyses.

At the galactic scales, the  results from pure dark matter simulations are known to be inaccurate because of the absence of baryonic feedback. It was shown by \citep{2011MNRAS.417.2020S, 2013arXiv1310.7571V, 2012MNRAS.420.2551B} that baryons and baryonic feedback can suppress the matter power spectrum from a few to 15-20 percents compared to a pure dark matter Universe. The power suppression affects all scales differently, depending on a particular combination of Active Galactic Nuclei, stellar winds and supernovae feedback. Mock galaxy catalogues constructed from dark-matter only simulations will therefore over predict the fluctuation amplitude at scales smaller than a megaparsec.
This is probably the main limitation of any  {purely} N-body {calculations}. In this work, we also ignore the effect of baryons and argue that the impact of finite volume effects can be revisited with another generation of hydrodynamical-simulations; likely the FS effect does not depend precisely on the fraction of matter that is baryonic and our correction model would still hold. This will be validated in a future study.

This paper is organized as follow. We start Section \ref{sec:theory} with a  contextual motivation for the measurement of uncertainty from weak lensing simulations. 
We then present the numerical setup of  a new suite of $N$-body simulations, the SLICS series (Scinet LIght Cone Simulations).
We  compare our  measurements of the two-point functions against predictions from three theoretical models.
Next, we briefly review the weak lensing theoretical background, we describe our light cone construction and present our estimators and models in Section \ref{sec:weaklensing}.  
In Section \ref{sec:structures}, we present general relations that describe the box effect as a function of redshift, angle and simulation box size.
 In Section \ref{sec:cov}, we turn our attention on  covariance matrices and examine how they are affected by the finite volume. 
We revisit the calibrations of the non-Gaussian $\xi_{+}$ covariance matrix\footnote{See Section \ref{subsec:xi_pm} for a definition of the $\xi_{\pm}$ quantities.} proposed by \citet{2007MNRAS.375L...6S} and \citet{2011ApJ...734...76S}
and propose an improved parameterization. 
To the best of our knowledge, we provide the first semi-analytical prescription to construct  the non-Gaussian estimator of the $\xi_{-}$ covariance matrix.
 In Section \ref{sec:application}, we discuss the practical implementation of the FS correction in data analyses pipelines,
in the presence of general source redshift distributions and survey masks.   
We conclude afterwards.

\section{Background}
\label{sec:theory}

This Section first reviews some of the challenges in the error estimation from galaxy survey,
then introduces the SLICS series and the set of theoretical models we use throughout the paper.  
We then present our light cone geometry and different estimators, 
and finally compare the measurements to theoretical predictions that incorporate the finite box effects.

\subsection{State of affairs}
\label{subsec:context}

Fourier modes corresponding to fluctuations larger than the simulation box size are also called `super'-modes. 
In a finite volume simulation, the missing super-modes inevitably affects, via non-linear mode coupling, the clustering properties of dark matter 
in real space \citep{2006MNRAS.370..691P} and in Fourier space \citep{2003MNRAS.341.1311S, 2008MNRAS.389.1675T,  2012JCAP...04..019D, 2013arXiv1304.7849H}.
In weak lensing as well, the missing super-modes propagate through the light cone,
as observed by \citet{2009ApJ...701..945S, 2011ApJ...734...76S}, which yields to a power and variance suppression over a large range of angular scales.

It is useful at this point to recall that there are other ways to measure covariance matrices than from an ensembles of mocks.
For instance, another approach is to estimate the error from a single large realization -- or from the  data itself -- by jackknife of bootstrap resampling sub-volumes. 
This is not an ideal approach, it was shown that  {\it internal} error such as bootstrap and jackknife estimates are biased by up to 40 per cent, 
due to the residual correlations between the sub-volumes \citep{2009MNRAS.396...19N}.
Moreover, this approach requires simulations with very large box size and high particle count, 
such as the MICE-GC  simulation \citep{2013arXiv1312.1707F}, in which the whole survey can  be fully embedded.
However, it is extremely challenging and expensive to run a Gpc simulation with a resolution 
as high as those required from weak lensing. The MICE-GC, for instance, shows $>$ 10 percent of loss in convergence power spectrum 
due to resolution limitations starting at $\theta = 27$ arc minute (or $\ell \sim 800$) for sources at $z_s=1$. 
These small scales are needed since the cosmic shear and weak lensing signals  pick contributions  both from linear and non-linear scales, 
with exact  proportions that depend on the estimator 
\citep[in][it is shown that the sensitivity of the three-dimensional weak lensing signal peaks at $k \sim 0.5\mathhMpc$ and spread significantly to $k$-modes well beyond unity]{2014arXiv1401.6842K}.


The precision requirements of current and future weak lensing surveys, where both small and large scales are important for scientific applications, 
seem to  {demand} an estimation of the covariance matrix from ensembles of realizations.
Running such large ensembles is computationally very expensive, and  generally requires a well thought-of tradeoff between the number of realizations $N_{sim}$, the cosmological volume $L_{box}$ and the resolution.
One of the results of this paper is that 
it seems possible to loosen our criteria on $L_{box}$.
A good resolution is of course crucial in order to preserve the non-linear signature of the signal within acceptable limits.
In addition, there has been a recent realization that a low $N_{sim}$ also has dramatic consequences.
As first pointed out by \citet{2007A&A...464..399H}, reducing the number of realizations inevitably leads to a noisy covariance matrix, and to a biased inverse matrix. It was shown by \citet{2013MNRAS.432.1928T} that a lower $N_{sim}$ leads to large error {\it on} the error bars, and that the number of simulations should be considerably more than the number of data points $N_{data}$. For instance, in order to reach a 5 percent error on cosmological parameters, one must have $N_{sim} = N_{data} + 200$ (see their equation 58).
The noise in the covariance matrix also contributes to an {\it additional} variance term on the cosmological parameters itself \citep{2013PhRvD..88f3537D}, which scales as $(1 + N_{data}/N_{sim})$. 
With the current simulations, we have $N_{sim} = 500$, hence the extra error would be less than 10 percent 
for a data vector of size $N_{data} < 50$. 
Given these constraints, the best strategy seems to reduce the size of the simulated volume while keeping both the resolution and $N_{sim}$ high.
It is then possible to correctly address and minimize the extra error term, at the price of allowing finite box effects to contaminate our calculations.
What matters then is to identify and keep track of all such effects, to correct those that can be corrected, and to account for the others
in the final error calculation.


\subsection{$N$-body simulations}
\label{nbody}

We construct convergence and shear maps from a large ensemble  of  $500$ $N$-body simulations -- the SLICS-LE --
with box size of $L_{box} = 505\mathMpch$ that are  based on the {\it WMAP9} + BAO + SN cosmology, namely:
$\Omega_m = 0.2905$, $\Omega_\Lambda = 0.7095$, $\Omega_b = 0.0473$, $h = 0.6898$, $\sigma_8 = 0.826$ and $n_s = 0.969$.
These follow the non-linear evolution of $1536^3$ particles inside a $3072^3$ grid cube, from $z_i = 120$ down to $z =0$.
The initial conditions are obtained from the Zel'dovich displacement of cell-centred particles, based on a transfer function obtained 
with the {\small CAMB} online tool \citep{Lewis:1999bs}.  The $N$-body calculations are performed with {\small CUBEP$^3$M} \citep{2013MNRAS.436..540H},
a fast and highly scalable public $N$-body code that solves Poisson equation on a two-level mesh,  and reaches sub-grid resolution from particle-particle interactions inside the finest mesh.
This code has been optimized for speed and minimal memory footprint, and is therefore well suited for such a task.
Each simulation was performed in about 30 hours  on $64$ nodes at the SciNet GPC cluster \citep{Scinet},
a system of IBM iDataPlex DX360M2 machines equipped with two Intel Xeon E5540 quad cores,  running at 2.53GHz with 2GB of RAM per core. 

At selected redshifts\footnote{We also produced dark matter halos at the same redshifts from the on-the-fly spherical overdensity halo finder  described in  \citet{2013MNRAS.436..540H}.
However, these haloes are not  part of the current paper, hence we leave their description for future work.} 
 and along each of the three cartesian axes, the particles are assigned on a $12,288^2$ cells grid following a `cloud in cell' (CIC) interpolation scheme \citep{1981csup.book.....H}. These  `mass planes' are stored to disks and serve in the construction of `lens planes' in the ray-tracing algorithm (see section \ref{subsec:lightcone}).
 Particles themselves are temporarily dumped to disk at $z = 0.640$ and $z = 0.042$ for dark matter power spectrum measurements, 
 after which the memory is released. 

As described in the {\small CUBEP$^3$M} reference paper, one of the limitation from the default configuration of this code is that the force calculation 
at the grid scale suffers from important scatter, which effectively smooths out some of the structure at scales up to 15 fine mesh cells. 
To quantify this effect and understand the resolution range on the SLICS-LE suite, we ran five simulations
in a high precision mode, in which the scatter in the force is minimized by extending  the exact particle-particle force 
up to two layers of fine mesh around each particles.
These `high resolution' simulations, the SLICS-HR suite, 
resolve scales about $4$ times smaller than the finer mesh. 

\subsection{ $P(k)$ - estimator}
\label{subsec:Pk}

The dark matter power spectrum $P(k)$ captures a large amount of cosmological information, and
 is related to the dark matter overdensity fields $\delta({\bf x})$ by : 

\begin{eqnarray}
\langle | \delta ({\bf k}) \delta ({\bf k'}) | \rangle = (2\pi)^{3}P({\bf k})\delta^3_{D}({\bf  k'} - {\bf k})
\label{eq:power}
\end{eqnarray}
with $\delta ({\bf k})$ the Fourier transform of $\delta({\bf x})$.
To measure this quantity from our simulations, we assign all the particles onto $3072^3$ cells, matching in resolution the finer mesh of the $N$-body code, 
and use the {\small FFTW}  libraries \citep{FFTW3} to perform the Fourier transform. 
Since it is computed on a grid, the power spectrum measurement is affected by the mass assignment scheme -- CIC in our case --  and we can partially undo this effect
with a simple procedure proposed by \citet{1981csup.book.....H}. i.e. by dividing the measured power spectrum
by the Fourier transform of the assignment scheme, i.e. : 

\begin{eqnarray}
P({\bf k}) = \frac{|\delta({\bf k})|^2}{H^4(k_x)H^4(k_y)H^4(k_z)} , \mbox{\hspace{1cm}} H(x) = \mbox{sinc}\bigg(\frac{\pi x}{nc}\bigg)
\label{eq:jing}
\end{eqnarray}

We could have opted for a more optimal deconvolution algorithm such as the iterative procedure proposed by \citet{2005ApJ...620..559J}, 
however we are mainly interested in weak lensing statistics, hence optimal accuracy of $P(k)$ at the grid scale is not necessary.
In the end, we estimate the isotropic power spectrum by taking the average over the solid angle: $P(k) = \langle P({\bf k}) \rangle_{\Omega}$.
Given the number of particles and volume probed, the contribution from shot noise is negligible over the scales that matter 
to us, hence we do not attempt to subtract it.

\subsection{ $P(k)$ - models}
\label{subsec:Pk_th}

At large scales (low $k$) the mass power spectrum is well described by Gaussian statistics. At small scales, non-linear mode coupling develop and different theoretical models are not in perfect agreement. The widely used {\small HALOFIT} model
 \citep{2003MNRAS.341.1311S} is missing 5-10 percent in the dark matter power spectrum between scales $0.1 < k < 1.0 h \mbox{Mpc}^{-1}$,
 and more than 50 percent for  $k > 10 h \mbox{Mpc}^{-1}$ \citep{2010ApJ...715..104H}. 
 It has then been subject to a recent recalibration  by  \citet{2012ApJ...761..152T}, which unfortunately seems to present instead
 an  overestimate of order five percent over the same range of scales for LCDM cosmology \citep{2013arXiv1304.7849H}. The original model -- \halofit2011, i.e before the 2012 recalibration --  still shows many advantages over the \halofit2012 model. 
Although it is less accurate at small scales, it is based on a larger suite of $N$-body simulation, hence its  dependence on cosmological parameters
is generally considered to be more finely calibrated, and is still widely used in likelihood analyses.
We therefore decided to consider both of these models for comparison with our $N$-body suites. 

One of the drawback of the \halofit \ approach is that it  attempts to describe the non-linear coupling with a single fitting function, 
and it is questionable whether this can truly capture all the information.
As an alternative, the Cosmic Emulator  \citep{2010ApJ...715..104H, 2013arXiv1304.7849H}
 is based on an interpolation between a set  of measurements from simulations, 
 which were performed  at  carefully selected points in parameter space.
In this paper, we thus consider this Cosmic Emulator as a third model, 
and  take advantage of the extended edition that achieves better than 5 percent precision on the 
dark matter power spectrum down to $k\sim10.0 h\mbox{Mpc}^{-1}$.
Unfortunately, the scope of the excursion in parameter space is not as large as other models, 
hence might not be adequate for some analyses. There have been efforts in the past to stitch 
the \halofit \ predictions on top of the Cosmic Emulator in order to cover cosmological parameter and $k$-modes
that are outside of the range of validity \citep{2011MNRAS.418..536E}, but this falls outside the scope of the current paper.

We finally note that the Cosmic Emulator is constructed for 37 cosmologies, each sampled with nested $N$-body simulations of different volumes.
The largest modes were assessed to be accurate to better than a percent by comparing results with simulations of $L_{box} = 2\mathGpch$.
In other words, the Cosmic Emulator was explicitly shown to be minimally affected by the finite box effects under consideration in the current work,
which ensures that the largest scales are fully reliable.  

\subsection{ $P(k)$ - results}
\label{subsec:Pk_comp}

\begin{figure}
   \centering
   \hspace{-1cm}
   \includegraphics[width=3.4in]{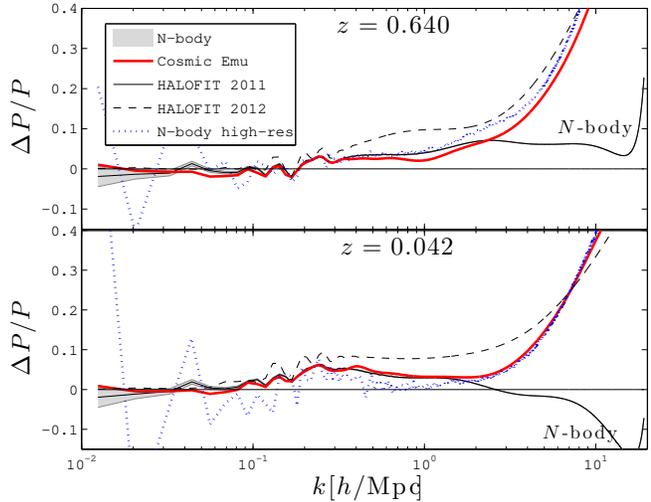} 
   \caption{Fractional error between various power spectrum measurements and the HF1 model, 
   at $z=0.640$ (top panel) and $z = 0.042$ (bottom panel).  
       The solid lines surrounded by shaded regions (labeled `$N$-body' at high $k$ values) represent the mean and $1\sigma$ error about the mean (i.e. $\sigma/\sqrt{N}$) 
       power from the SLICS-LE simulation suite. 
           The blue dotted lines with large scatter represent measurements from the SLICS-HR suite. 
    Also shown are results from the HF2  and CE predictions. 
    The HR and CE measurements exhibit the highest level of agreement. 
}
   \label{fig:3dpk}
\end{figure}

Fig. \ref{fig:3dpk} shows the power spectrum measured in the SLICS-LE and -HR suites at $z = 0.640$ and $0.042$, and compared  to the three prediction models. 
All models and measurements are relatively close to each other, therefore we focus on the fractional difference with respect to  {\small HALOFIT}2011 (HF1 hereafter). 
We adopt this convention throughout the paper unless otherwise specified.

Up to $k = 2.0 h \mbox{Mpc}^{-1}$, the SLICS-LE and -HR simulations suites 
match the Cosmic Emulator (CE hereafter) predictions to within 
two per cent, whereas it deviates from the other two \halofit \ models by up to 6 percent. 
The HR simulations present significant scatter at the largest scales, as expected when dealing with only a handful of realizations. 
For the LE sample, the mean from the  {largest scales (low $k$)} is biased low  by about two per cent, 
as expected from the incomplete capture of the linear regime in a finite box environment smaller than 1Gpc \citep{2008MNRAS.389.1675T}.
This is fully consistent with the results  of \citet{2010ApJ...715..104H}.

Beyond $k=2.0h\mbox{Mpc}^{-1}$, the LE simulations suite lacks power compared to CE and HF2, whereas it is in reasonably good agreement with HF1.
 {This} is caused by a resolution loss, as shown by the HR simulations. HR closely  follows CE with a maximum deviation of 3 percent up to 
$k=10.0h\mbox{Mpc}^{-1}$. At these smallest scales however, HF2  differs  by at least 5 percent with respect to CE, on the low or high side depending on redshift. 
From these results, we reach the following conclusions
\begin{enumerate}
\item Scales smaller than $k=2.0h\mbox{Mpc}^{-1}$ in the LE suite  are affected by the resolution limits of the $N$-body code 
\item  {HF2 over-predicts the HR power  by more than 5 percent in the range $0.5 < k < 1.5 \mathhMpc$ at $z = 0.64$, and in the range 
$0.5 < k < 4 \mathhMpc$ at $z = 0.042$,}
\item The CE model provides the best fit to our HR simulations up to its limit at $k=10\mathhMpc$, and
\item Box effects are  affecting our $P(k)$ measurement by no more than 2-3 percent, at the largest scale only.
\end{enumerate}
Keeping this in mind, let us now turn our attention to the light cone measurements, where these results  propagate in $\ell$-space.


\section{Weak Lensing}
\label{sec:weaklensing}

\subsection{Light cone construction}
\label{subsec:lightcone}

Weak lensing simulations are generally constructed by integrating over null geodesics 
 in the past light cone, using the full volume \citep{1999MNRAS.308..180C}
 or a set of discrete lens planes \citep{2002ApJ...570...17M}.
The  integration  can either be computed along photons trajectories \citep{2003ApJ...592..699V}
or carried on straight lines under Born's approximation. Differences between these techniques
are small and occur mainly at the smallest scales, hence have no consequence on our results \citep{2002ApJ...574...19C, 2003PhRvD..68h3002H}.
For simplicity, we work with multiple lens planes, using the Born approximation and assuming flat sky.

In the thin lens approximation, the weak lensing convergence $\kappa({\boldsymbol \theta})$ can be obtained by integrating over
the density contrast $\delta(\chi,{\boldsymbol \theta})$ and the source distribution $n(z)$ along the line of sight:
\begin{eqnarray}
 \kappa({\boldsymbol  \theta} ) = \frac{3 H_{0}^{2} \Omega_{m}}{2 c^2} \int_{0}^{\chi_H} \delta(\chi,{\boldsymbol \theta}) (1 + z)  g(\chi) \chi  d\chi 
 \label{eq:kappa}
 \end{eqnarray}
Here,  $H_0$ is the Hubble constant, $c$ the speed of light, $\chi_H$  the comoving distance to the horizon and $g(\chi)$
is related to the source distribution via:
 \begin{eqnarray}
         g(\chi) =  \int_{\chi}^{\chi_H} n(\chi') \frac{\chi' - \chi}{\chi'} d\chi' 
\end{eqnarray}
The mapping between $n(z)$ and $n(\chi)$ involves the Jacobian: $\mbox{d}\chi / \mbox{d} z = c/H(z)$ and
$n(\chi)$ is normalized as $\int n(\chi) d\chi = 1$. 
In this paper, we focus on  the case where the sources are assigned at a fixed redshift $z_s$, i.e. $n(z) = \delta_{D}(z - z_s)$, 
and discuss the more realistic scenario of a broad redshift distribution in Section \ref{subsec:nz}. 

Two-dimensional density fluctuations planes are constructed from collapsed density fields, i.e. $\delta_{2D}({\boldsymbol \theta}) = \sum_{\chi} \delta(\chi,{\boldsymbol \theta}) d\chi$, where $\chi$ is the coordinate along one of the cartesian axis.
This effectively turns the integration along the photon trajectory into a discrete sum at the lens locations: $\int_{0}^{\chi_H} d\chi \rightarrow \sum_{\chi_{lens}} \Delta \chi_{lens}$.
In our post-processing ray-tracing code, we construct  light cones by integrating the lens planes from the observer to $\chi_{max}$,
and by interpolating linearly  onto $6000^2$ pixels spanning a fixed opening angle of about 8 degrees on the side, thus defining our light cone lens planes $\delta_{2D}(\chi_{lens},{\boldsymbol \theta}_{pix})$.
Following the pioneering work of \citet{2002ApJ...570...17M}, we use the common approach of random shifting and rotating each mid-planes  to minimize undesired correlations between the different lens planes.
As described in \citet{2012MNRAS.426.1262H}, we compute the shear maps $\gamma_{1,2}$ from the convergence via their common coupling to the potential field:
\begin{eqnarray}
\kappa = \frac{\Phi_{,xx} + \Phi_{,yy}}{2}, \mbox{\hspace{1cm}} \gamma_1 = \frac{\Phi_{,xx} - \Phi_{,yy}}{2},  \mbox{\hspace{1cm}}  \gamma_2 = \Phi_{,xy}
\end{eqnarray}
To avoid boundary effects and numerical noise, we solve these equations with Fourier transforms of the full periodic mass maps $\delta_{2D}({\boldsymbol \theta})$.
The interpolation on the pixels is done at the very end, on the shear and convergence maps all at once.

The broad redshift distribution of the sources is also binned, which turns the integral over the sources into a sum:
$\int_{\chi}^{\chi_H} d\chi' \rightarrow \sum_{\chi_s = \chi_{lens}}^{\chi_H} \Delta \chi_s$. 
Inserting both pieces into equation \ref{eq:kappa}, we obtain our discrete equation:
\begin{eqnarray}
 \kappa( {\boldsymbol \theta}_{pix} ) = \frac{3 H_{0}^{2} \Omega_{m}}{2 c^2} \sum_{\chi_{lens}=0}^{\chi_H} \delta_{2D}(\chi_{lens},{\boldsymbol \theta}_{pix}) (1 + z_{lens})  \chi_{lens} \times \nonumber \\
         \bigg[ \sum_{\chi_s = \chi_{lens}}^{\chi_H} n(\chi_s)\frac{\chi_s - \chi_{lens}}{\chi_s}  \bigg] d\chi_{lens} d\chi_s
         \label{eq:kappa_disc}
\end{eqnarray}

The choice of $z_{max} = 3.0$ -- the farthest lens plane -- is set to be larger than the maximum source redshift of current and future weak lensing surveys.
The opening angle of the light cone is set to $60$ square degrees, 
which spans 10 percent of the simulation box at $z = 0.13$, 50 percent at $z = 0.75$, and matches the box at $z = 2.0$. 
We allow the light cone to extend up to $z=3$ by using periodic boundary conditions to fill in regions of the lensing maps  
that otherwise fall outside the simulation volume. These redshifts $z>2$, where the box is repeated, are strongly affected by the missing large scale super $k$-modes, 
but for a broad redshift distribution, the contribution to the weak lensing signal coming the high redshift tail is  {minimal}.
It is therefore generally accepted that these repetition effects have a negligible role to play in the systematics budget of the mock catalogues.
With the inclusion of the FS effect described in this paper, this statement can finally be made accurate (see Section \ref{subsec:trends}).

With the increasing importance of three dimensional and tomographic weak lensing analyses, 
is has become clear that a fine redshift sampling is essential in order to capture accurately the growth of structures.
With the adopted cosmology and $L_{box}$, stacking $9$ simulations cubes back-to-back continuously fills the space
up to  $z=3$, which leaves very little prospect to calibrate tomographic analyses on these simulations. 
We double the redshift sampling by projecting only half the simulation box,  i.e. 
volumes that are  $257.5  \mbox{Mpc}h^{-1}$ thick, in the construction of  each of the 18 lens planes. 
 For example, the first lens is produced by collapsing a volume whose front end is at the observer and far end at $\chi=L_{box}/2  = 257.5 \mbox{Mpc}h^{-1}$.
  This volume is assigned to its central comoving position, i.e. at $126.25  \mbox{Mpc}h^{-1}$,
which corresponds to $z = 0.042$. The second lens is collapsed at the centre of the adjacent half box, i.e.   at $378.75 \mbox{Mpc}h^{-1}$,
corresponding to $z = 0.130$, and so on. Generally, lens planes are generated at $\chi_{lens} = [(2n-1)/4] L_{box}, n = 1,2,...$

With this setup, what we call the set of `natural' source planes are those located at the rear faces of each collapsed volumes, i.e at $ \chi_s = \chi_{lens} = [n/2] L_{box}, n = 1,2,...$.
These source planes are special as they can be used in equation \ref{eq:kappa_disc} without any interpolation in redshift. 
Otherwise, a measurement from a  general source plane will receive contributions from a fraction of a lens, 
which we calculate from an interpolation between the enclosing natural source planes.
The $18$ lens and natural source planes are summarized in Table \ref{table:redshifts}.

At the lowest redshifts, only a tiny fraction of the mass plane is used in the ray-tracing algorithm.
It is tempting to recycle some of these volumes for more than one light cone, but this comes at the cost of inducing an extra level of correlation 
in the covariance matrix. Since this is exactly what we want to measure, we avoid this situation in the LE suite 
and work exclusively with independent realizations. 
The situation is different for  the HR suite, which is not directly used for covariance matrix calculations, but rather for checking the convergence
of the small scales. We therefore run only one of these costly simulation until the last lens redshift of $z = 0.042$, and stop the four companions  at $z = 0.221$.
For the construction   
 {of the HR light cones, we thus used 5 independent simulations for all lens planes with $z\ge0.221$, each completed with 
  a distinct, unique, region of the $z=0.130$ and $z=0.042$ mass planes extracted from a single simulation. }.

 \begin{table*}
   \centering
   \caption{Redshifts of the lens  planes and natural source planes that enter equation \ref{eq:kappa_disc}. 
   These are obtained by stacking half boxes, each $257.5 h^{-1}\mbox{Mpc}$ thick, 
   from the observer to $z_{max} \sim 3.0$, given the fiducial cosmology.
   }
   \tabcolsep=0.11cm
      \begin{tabular}{@{} lllllllllllllllllll @{}} 
      \hline
    $z_{lens}$ & 0.042 &  0.130 &   0.221    &0.317 &  0.418 &   0.525    &0.640 &  0.764    &0.897  &  1.041  &  1.199    &1.373  &  1.562  &  1.772  &   2.007   & 2.269 &   2.565    &2.899\\
  \hline
  $z_{s}$& 0.086  &  0.175  & 0.268 &   0.366   & 0.471    &0.582  &  0.701   & 0.829  &  0.968    &1.118   & 1.283  &  1.464   & 1.664 &   1.886   &  2.134 &   2.412   & 2.727 &   3.084\\
 \hline
   \end{tabular}
    \label{table:redshifts}
    \end{table*}

\subsection{ $C^{\kappa}_{\ell}$ - estimator}
\label{subsec:C_ell}

We compute the weak lensing power spectrum $C_{\ell}^{\kappa}$  from each simulated map with two-dimensional fast Fourier transforms\footnote{
Following the three dimensional case, we minimize the impact of the grid assignment scheme with a deconvolution of the $C_{\ell}$.
In this case, we apply equation \ref{eq:jing} in two-dimensions.}, assuming a sky flat, 
and average over annuli that are linearly spaced in $\ell$ . 
This approach to the measurement suffers from some systematic effects which are worth mentioning.
Firstly, we do not take into account the non-periodic nature of the light cone, which introduces non-isotropic features in the Fourier transformed map.
However, this effect is small to start with,  and is further suppressed to a sub-percent level effect during the angle averaging.
Secondly, the  {strong} interpolation  {inherent to the pixelization} of the lower redshift planes introduce an artificial smoothing effect in addition to the intrinsic softening length of the $N$-body code.
Although these planes  are strongly  down-weighted  with deep lensing surveys, they systematically reduce the signal compared to a full particle light cone. 
Finally, even though correlations between different lenses
are reduced by random rotations and origin shifting, the residual correlations translate into cross-terms
in the calculation of $C_{\ell}^{\kappa}$, which can be of a few percent.

\subsection{ $C^{\kappa}_{\ell}$ - models}
\label{subsec:C_ell_th}

The weak lensing power spectrum $C_{\ell}^{\kappa}$ is computed using the Limber approximation \citep{1954ApJ...119..655L} to perform the line-of-sight integration over $P(k,z)$:
\begin{eqnarray}
C_{\ell}^{\kappa} = \int_{0}^{\chi_{\infty}} d\chi \frac{W^2(\chi)}{\chi^2} P(\ell/\chi,z) = \frac{1}{\ell}\int_{0}^{\infty} dk W^2(\ell/k) P(k,z) 
\label{eq:limber}
\end{eqnarray}
where $\ell = \chi k$, $z = z(\chi) = z(\ell/k)$, and the lensing kernel $W(\chi)$ is given by
\begin{eqnarray}
W(\chi) = \frac{3 H_{0}^{2} \Omega_{m}}{2 c^2}\chi g(\chi)  (1 + z)
\end{eqnarray}
We use  $1 \le \ell \le1\times 10^5$, where high $\ell$ values are necessary when working in real space and smoothing windows (see Section \ref{subsec:xi_pm}). The truncation of  {super-modes} is computed following \citet{2011ApJ...734...76S} {as a low $k$ cut}. 
We also investigate the effect of truncating scales smaller than the intrinsic CE cutoff at $k \sim 10.0 h\mbox{Mpc}^{-1}$, {in the form of high $k$ cut},   in order
to fully understand the consequences of the cut at small scales. Low and high $k$ cuts are included in the predictions using equation \ref{eq:limber}. The different models, six in total, are summarized in Table \ref{table:models}. 

\subsection{ $C^{\kappa}_{\ell}$ - results}
\label{subsec:C_ell_comp}

\begin{table}
   \centering
   \caption{Different prediction models considered in this paper. In most of the calculations, we use $L_{box}$  = $505  h^{-1}\mbox{Mpc}$ (or $k_{box} = 0.0124 \mathhMpc$).
   For the Cosmic Emulator predictions only, the box size is allowed to vary to $1000$, $505$, $257$ and $147$ \Mpch. 
   We regroup all these volumes under the quantity $L_{var}$, as  indicated in this Table. 
   Note that both the CE and CEk models have a small scale $k$ cut at $10.0 \mathhMpc$, which is the resolution limit of the Cosmic Emulator.}
      \begin{tabular}{@{} ll ll } 
      \hline
Model &   $k$-modes included  & Name\\
             & [ in $h \mbox{Mpc}^{-1}$]  & \\
\hline
           {\small HALOFIT}2011                       &    $0.0010 < k < 40.0$                 &  HF1\\
           {\small HALOFIT}2011 + $k$-cuts   &    $0.0124  <  k < 10.0  $     &  HF1k\\
           {\small HALOFIT}2012                      &     $0.0010 < k < 40.0$                 &  HF2\\    
           {\small HALOFIT}2012 + $k$-cuts   &    $0.0124 <  k < 10.0$       &   HF2k\\
           Cosmic Emulator                                   & $ 0.0010 < k < 10.0$            &  CE\\    
           Cosmic Emulator  + $k$-cuts              & $ 2\pi/L_{var} <  k < 10.0 $ & CEk  \\    
      \hline
   \end{tabular}
    \label{table:models}
\end{table}

\begin{figure}
   \centering
     \includegraphics[width=3.4in]{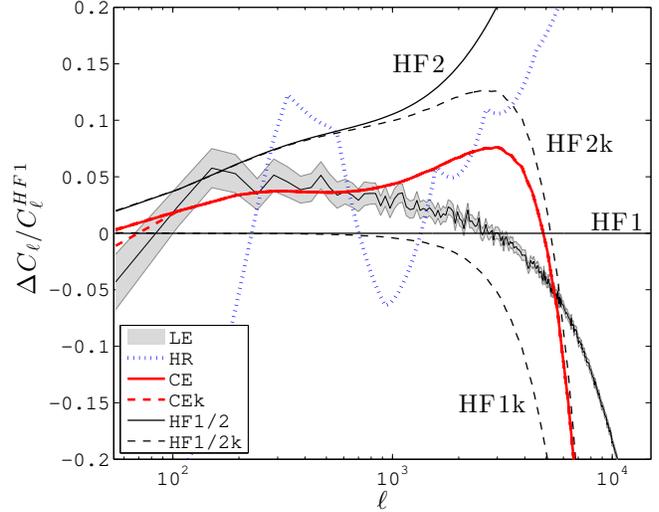} 
   \caption{Fractional error between various weak lensing power spectrum measurements and the HF1 predictions, at $z_s = 0.582$. 
   We show results from the two simulations suites (LE and HR) plus all models of Table \ref{table:models}.
   The cut at $k = 0.0124 \mathhMpc$ impacts scales with $\ell< 60$ for the source redshift considered here, hence is not visible in this figure.
   For the different values of $L_{var}$ considered with the CEk model, only the cut at $k =2\pi/147 \mathhMpc$ 
   is visible,  shown with the (red) thick dashed line. 
 The small scale $k$ cut at $k = 10.0 \mathhMpc$, common to the HF1k, HF2k, CE and CEk models, impacts all measurements  
 at $\ell> 1000$.
   Results from other redshifts are qualitatively similar.
   }
   \label{fig:l2Cl}
\end{figure}

Fig. \ref{fig:l2Cl} shows the weak lensing power spectrum for the six models of Table \ref{table:models} and the sources placed at the `natural' slice at redshift $z_s = 0.582$. Other redshifts are qualitatively similar.
Results are presented in terms of fractional error with respect to the  HF1 model.
As first pointed out in \citet{2012ApJ...761..152T}, the {\small HALOFIT}2012 calculations depart by more than 10 per cent  with respect to {\small HALOFIT}2011 for $\ell> 1000$.
Because weak lensing measurements project many scales onto each angle, the two models match only at the lowest multipoles. 
The CE lies roughly halfway in between HF1 and HF2,
and features a sharp cutoff at $\ell \sim 3000$ that is caused by the exclusion of small scale modes with  $k> 10.0 h\mbox{Mpc}^{-1}$.
The same high $k$ cut applied on the {\small HALOFIT} models, HF1k and HF2k, shows a deviation from the no $k$ cut case for $\ell > 1000$. 
 {Consequently}, one can deduce that the CE model with no high $k$ cut would be higher by about $5$ percent at $\ell=3000$ and 
{most likely would} then 
follow the high resolution curve, HR, for higher $\ell$.

With $L_{box} = 505\mathMpch$ and the {multipoles} considered,  the volume effects are negligible.
We expect the effect of finite box size to be enhanced at higher redshifts, lower $\ell$ and for smaller simulation boxes.
To explore this, we vary $L_{box}$ in the CEk predictions,  and consider box sizes of $1000$, $257$ and $147h^{-1} \mbox{Mpc}$.
Figure \ref{fig:l2Cl} shows that, for this source redshift distribution, the power spectrum is not affected except for the lowest multipole at $\ell = 45$ and the box size $L_{box} = 147\mathMpch$ (CEk shown by the red dashed line). All other box sizes are indistinguishable from the CE model.
In general, the signal at low $\ell$ is affected only when probing the density fluctuations at angles where the light cone actually extends beyond the simulation box. For our setup this occurs at $z=2$. The SLICS-LE suite and HF1 are consistent within 5 percent for  $\ell<6000$, and are systematically lower than the HF2 model at all scales. As mentioned in the last section, this is an expected behaviour due to the limited resolution of both HF1 and SLICS-LE.

As expected from the $P(k)$ comparison in Section 2.5, the CE $C_\ell^\kappa$ provides the best
match to SLICS-LE over the scales that are resolved, that is within $5$ percent for $ \ell<2000$.

\begin{figure*}
   \centering
   \includegraphics[width=3.2in]{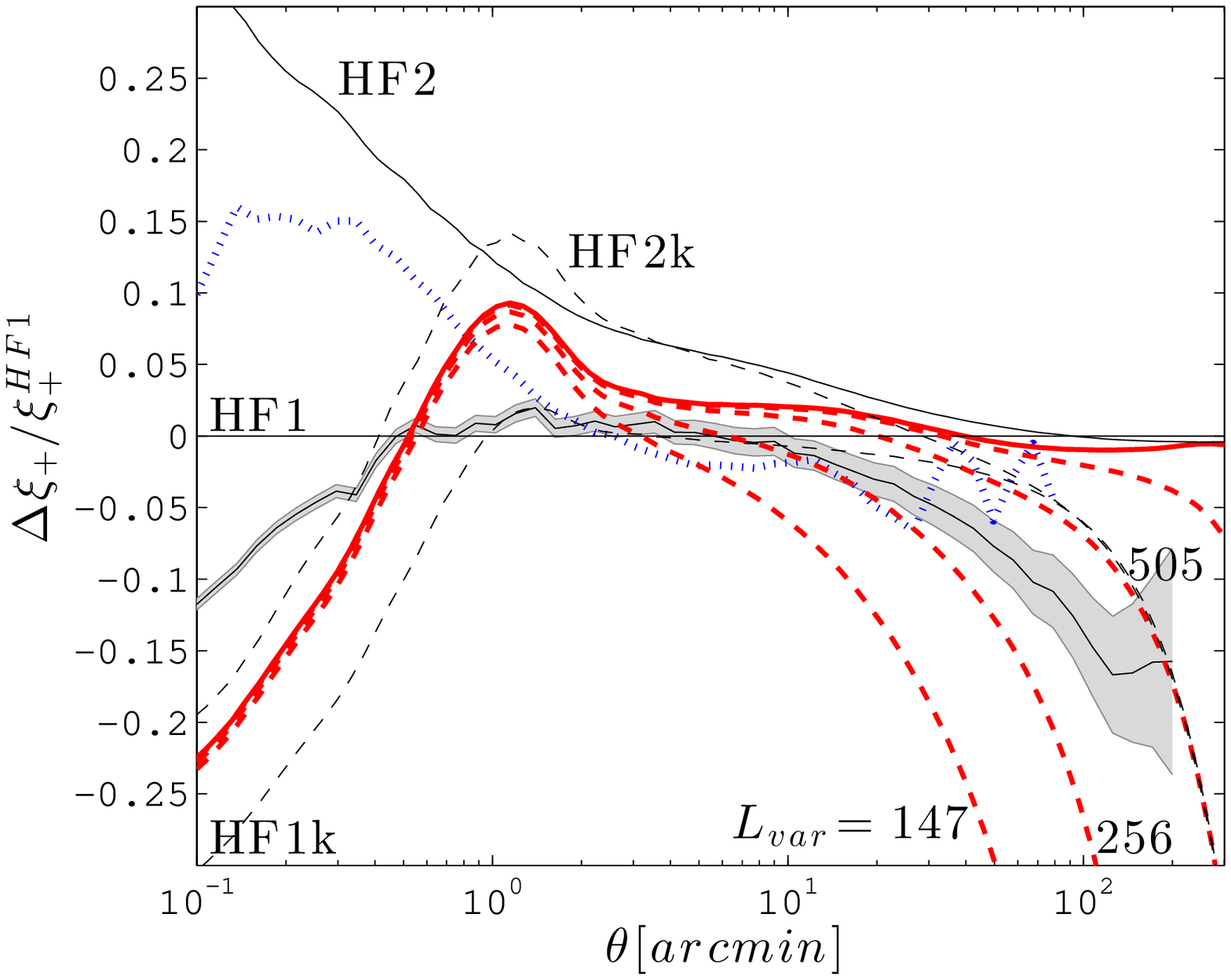}
   \includegraphics[width=3.2in]{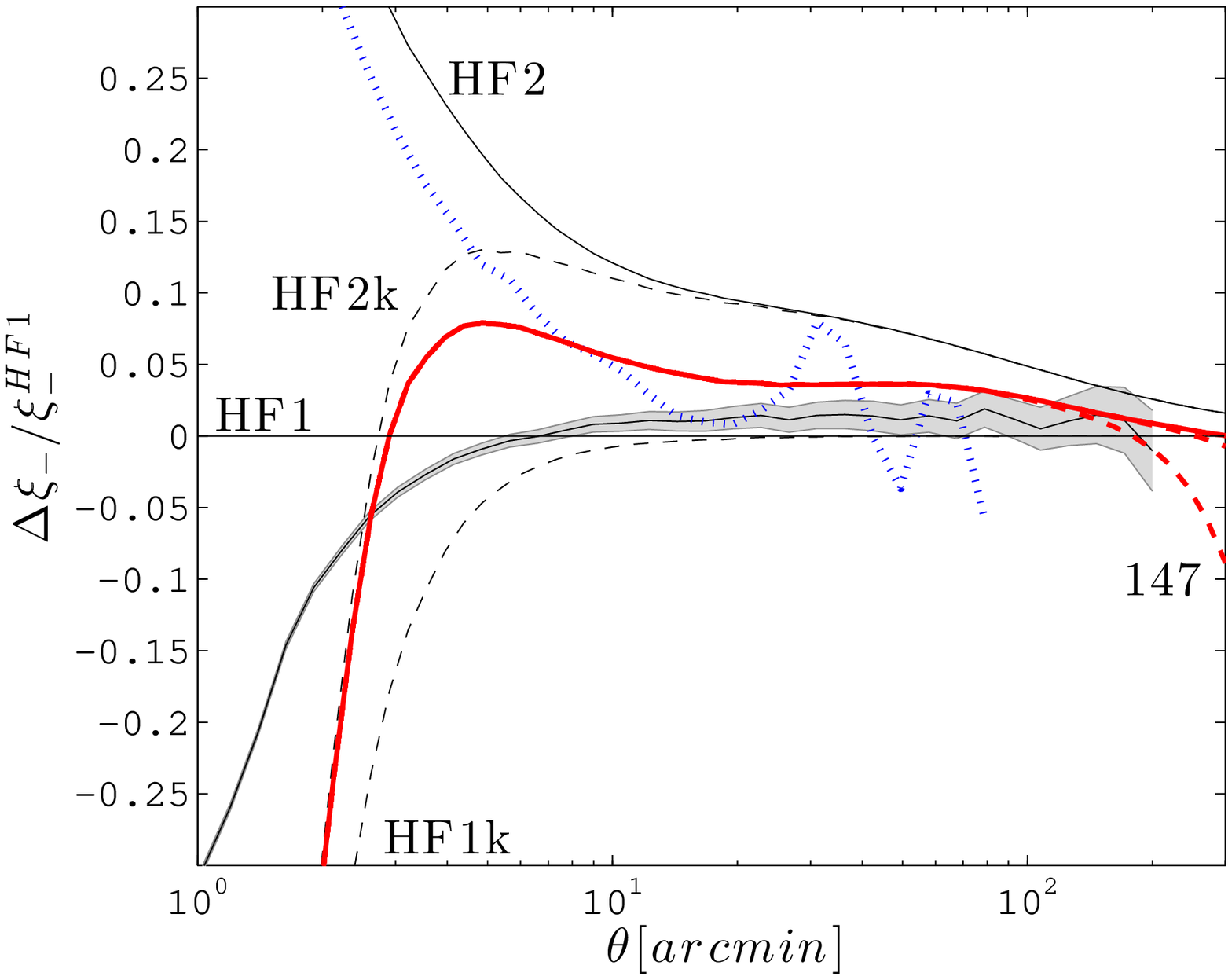}
   \caption{Fractional error on $\xi_{+}$ (left) and $\xi_{-}$ (right) with respect to the {\small HALOFIT}2011 predictions, for $z_s = 0.582$.  
   We show results from the two simulations suites (LE and HR) plus all models of Table \ref{table:models}.
   The cut at $k = 0.0124 \mathhMpc$ becomes important in $\xi_+$, and all models  converge to the same level of suppression (indicated by the label `505' in the figure). 
   For the CEk model, four different values of $L_{var}$ are now visible in the left panel -- $147$, $256$, $505$ and $1000 \mathMpch$, while only $L_{var} = 147 \mathMpch$ 
   is seen in the right panel. 
  We note that the models disagree significantly  at scales close to the arc minute. 
   }
   \label{fig:xipm}
\end{figure*}

\subsection{$\xi_{\pm}(\theta)$ - models}
\label{subsec:xi_pm}

Many gravitational lensing estimators are defined in real space. Any second order cosmic shear statistics, real space or not, can be expressed as a linear combination of $C_{\ell}^{\kappa}$ via an integral over $\ell$, weighted by a 'response function' in $\ell$-space. Without any loss of generality, we focus exclusively on the $\xi_{\pm}$ quantities, which are given by:
\begin{eqnarray}
    \xi_{\pm}(\theta) = \frac{1}{2\pi}\!\int\!\!\!C_{\ell}^{\kappa} J_{0/4}(\ell\theta)    \ell d\ell = \frac{1}{2 \pi} \!\int \!\!\!\! \int \!\!\! J_{0/4}(\ell\theta)\! W^2(\ell/k) P(k) d\ell dk.
    \label{eq:xi}
\end{eqnarray}
Note that we expect $\xi_{+}(\theta)$ to be more affected by the finite box effect than $\xi_{-}(\theta)$ because of $J_{0}(x)$ which peaks at small $x$. 

For each model listed in Table \ref{table:models}, we use equation \ref{eq:xi} to compute the $\xi_{\pm}(\theta)$ theoretical predictions. The results are shown in
Fig. \ref{fig:xipm} for $z_{s} = 0.582$. The different $\xi_{+}(\theta)$ {models} (left panel) with no $k$ cuts agree within less than $5$ percent at scale larger than $10$ arcmin.  
The cutoff of modes larger than  $k= 10.0 h \mbox{Mpc}^{-1}$ results in a sharp turn over at scales just under the arc minute for all models.
This suggests that a $\xi_{+}$ analysis based on scales larger than one arc-minute using the current Cosmic Emulator as an alternative to {\small HALOFIT} would be accurate within a few percent.
The HF2 predictions is consistently higher than other models, and depart by about $3-5$ per cent from the CE in the range $1 < \theta < 10'$. 

It appears that cutting out the small scale $k$-modes produces a small bump at the turnover scale of $1'$, 
as seen in the two {\small HALOFIT} predictions. This is a numerical effect due to the sharp $k$ cutoff and the oscillatory $J_0$ function.
It is therefore likely that in absence of the built-in small scale $k$ cuts, the CE model would be slightly lower, by about two percent, at the turnaround scale. 
The differences between the models below one arc-minute are large, and this source of theoretical uncertainty will have to be addressed in weak lensing analyses involving these scales. Note that this is also a scale where the baryonic effects are also very important.

A striking feature seen in the left panel of Fig. \ref{fig:xipm} is the impact of the large scale, low $k$, cut.
In all three models, excluding modes larger than the $505 \mathMpch$  box  produces significant suppression of power,
of one per cent for $\theta \sim 10'$, $5$ percent by a degree, and by more than $10$ per cent for $\theta > 2$ degrees. 
The set of thick red dashed lines in the figure shows CEk predictions for different box sizes. As expected, the smaller the box the larger the effect. Even the $1 \mbox{Gpc} h^{-1}$ box suffers from a 5 percent power loss at $\theta = 2^{o}$.

The $\xi_{-}$ signal (right panel) is by construction sensitive to smaller scales. This explains why the differences between models for scale below $10$ arc minutes is significantly larger than for $\xi_{+}$; at small scale, the different models with different high $k$ cuts are very different. On the other hand, the finite box effect is small, invisible for $\theta<100'$, it is only a few percent at $\theta=200'$. 
One needs to probe much larger angles in order to see the effects of the missing super modes.
When that occurs, however, the signal drop is very steep, and angles larger than a few degrees 
deviate from the `no $k$-cut' models rapidly. 

The $N$-body simulations used for the CFHTLenS mock catalogue, described in  \citet{2012MNRAS.426.1262H} and \citet{2012MNRAS.427..146H}, were partly  based on volumes as small as $147 \mbox{Mpc}h^{-1}$. It was not known at that time that the box size effects had the strong impact shown on the left panel of Fig. \ref{fig:xipm}, although departures from predictions were indeed observed. In the construction of the CFHT lensing covariance matrices,
large scale elements were stitched to  linear predictions to compensate for this effect \citep{2013MNRAS.430.2200K}.
We come back to this topic in Section \ref{sec:application}.

\subsection{$\xi_{\pm}(\theta)$ - estimator}
\label{subsec:xi_pm_est}

The shear correlation function $\xi_{\pm}$ from the SLIC simulations are measured from $500,000$ randomly sampled points on the shear maps. The LE simulation measurements shown in Fig. \ref{fig:xipm} agree within $5$ percent with HF1 and CE for $0.2 < \theta < 20'$. High and low $k$ cuts {built} in the simulation explain the loss of power outside this range of scale. HF2 is significantly higher than the simulations and the other models for all $\theta$.
This behaviour is consistent with the $P(k)$ and $C^{\kappa}_{\ell}$ comparison discussed in the previous sections. 
The high resolution suite exhibits a large scatter for $\theta>20'$, caused by the sampling variance, 
but it is in agreement with the LE suite, CE and HF1 at intermediate angles; at sub-arc minute scales, 
HR is systematically $\sim 10$ percent below HF2.

The low $k$ cut explains the large angle suppression since it is mixed to a wide range of angles when seen in projection; at large separation angle, the simulations agree with all the predictions that include the large scale $k$ cuts. 
The fact that the measurement fall between the $257$ and the $505 \mathMpch$ box predictions at large angles -- 
{  rather than exactly on top of the latter} -- suggests 
that there are residual systematics either in our modelling or in the measurement.
 It is difficult to pin down exactly what causes this small discrepancy, but 
it is no more than five percent over all angles. 
There is no evidence of finite box effect in $\xi_-$ {for the LE suite}, as expected from the theoretical calculations. The agreement with HF1 and CE is within $3$ percent for a separation angle larger than $10'$. The small scale signal is strongly affected by the finite resolution of the simulations, as exemplified by the comparison with the HR simulations. 

\section{Emerging structure in finite box effects}
\label{sec:structures}

We have shown in the last Section how a finite simulation volume affects weak lensing measurements as a function of the box size and the measurement angle/multipole. 
In addition, the FS effect also {varies with}  redshift, depending on what fraction of the simulation box at redshift $z_s$ is covered by the light cone. 
In this Section, we investigate quantitatively the universality of the FS suppression as a function of $(L_{box}, z_s, \theta,\ell)$, and propose a simple semi-analytical recipe to model its effect.

\subsection{$\ell$-space}
\label{subsec:trends}

\begin{figure}
   \centering
    \includegraphics[width=2.5in]{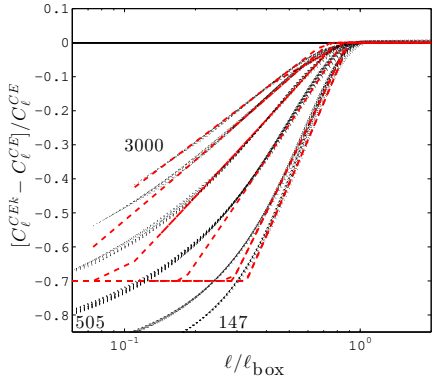} 
   \caption{Fractional error between the CE and CEk models, for box sizes 
    of $L_{var} = 3000$, $2000$, $1000$, $505$, $256$ and $147 \mathMpch$
  (we have labeled only a few of these in the figure for clarity). %
  We stack in this figure the measurements for $z_s = 0.582$,  $0.701$, $0.829$, $0.968$, $1.118$, $1.283$, $1.464$, 
  $1.664$, $1.886$, $2.134$, $2.412$, $2.727$ and $3.084$, and plot each of them versus $\ell/\ell_{box}(z)$ (see main text for a definition of $\ell_{box}$). 
  All these measurements are shown with the thin dotted lines, which superimpose remarkably well. 
     The $L_{box} = 505\mathMpch$ lines include the same calculations, carried this time with the HF2 and HF2k models. 
     These are not distinguishable from the CE/CEk calculations, which demonstrates that the results from this Figure are model independent.
    The thick dashed lines (red in the online version) represent the truncated power law fits, as defined in equation \ref{eq:f_ell}.    
     }
   \label{fig:l2Cl_fracbox_angles}
\end{figure}

{Following equation \ref{eq:limber}, the exact amount of weak lensing power lost by cutting out low-$k$ modes is given by:
\begin{eqnarray}
  C_{\ell}^{lost}(k_{box}) =   \frac{1}{\ell}\int_{0}^{k_{box}} \!\!\!\!\!\!\! dk W^2(\ell/k) P(k,z) = \int_{\ell/k_{box}}^{\infty} \!\!\!\!\!\! d\chi \frac{W^2(\chi)}{\chi^2}P(\ell/\chi,z)
  \label{eq:C_ell_lost}
\end{eqnarray}
with $k_{box} = 2\pi/L_{box}$. 
Let us inspect these equations and replace the higher bound of the $\chi$ integration by the distance to the furthest source,  $\chi(z_s)$.
Since $k_{box}$ is small and $W(\chi) = 0$ for $\chi \ge \chi(z_s)$, we recover from these expressions that only the lowest values of $\ell$
produce a non-zero integrand. 
We also read that for a given $\ell$, the integration picks up a narrow contribution ranging from $\ell/k_{box}$ to $\chi(z_s)$; 
closer sources and larger multipole rapidly shrink the integration range. 
For example, in our geometrical setup, the light cone exits the simulation box at $z = 2.0$. 
With $k_{box} = 0.0124 \mathhMpc$ and $z_s = 2$, we find that $C_{\ell}^{lost}  > 0$ for $\ell < 45.5$.
 At $z_s = 3$, $C_{\ell}^{lost}  > 0$ for $\ell < 55.9$.
 
{It is important to realize from equation \ref{eq:C_ell_lost} that the lower bound  of the integration 
makes the power loss a strong function of the ratio $\ell/k_{box}$.
Let us define $\theta_{box}(z_s)$ as the angle on the sky subtended by $L_{box}$ at redshift $z_s$.
Then we have $ \ell_{box}(z_s) = 2\pi/\theta_{box}(z_s) - 1= \chi(z_s) k_{box} $, 
and the ratio $\ell/k_{box}$ becomes $\chi(z_s)\times[\ell/\ell_{box}(z_s)]$ \footnote{It is important not to confuse $L_{box}$ (the size of the simulation volume, in units of Mpc/$h$) with $\ell_{box}$ (the multipole corresponding to an object of size $L_{box}$ on the sky, which is a redshift dependent quantity). }. 
Although the integral is challenging to solve exactly due to the complexe redshift dependences,
it is easily performed numerically with most Limber integration codes. 
The dominant dependence of $C_{\ell}^{lost}$ on $\ell/\ell_{box}(z_s)$ naturally emerges, as we explore in the remaining of this Section.  


\begin{figure}
   \centering
    \includegraphics[width=3.0in]{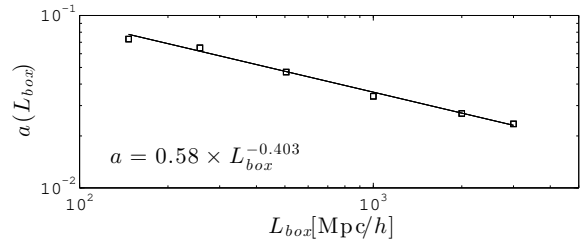} 
   \caption{Scaling of the power loss in $C_{\ell}^{\kappa}$, parameterized by $a(L_{box})$,  as defined in equation \ref{eq:f_ell}. 
   Open symbols are measurements  from the Cosmic Emulator calculations with six different box sizes, shown in Fig. \ref{fig:l2Cl_fracbox_angles}.
   The straight line is the best fit to the scaling  relation. }
      \label{fig:a_lbox}
\end{figure}

Fig. \ref{fig:l2Cl_fracbox_angles} shows the difference between the $C_\ell$ for the CE and CEk models relative to CE, as a function of $\ell / \ell_{box}$. We include predictions for the four box sizes  listed in Table \ref{table:models} plus two larger boxes, of $2$ and $3 \mathGpch$.
The results are also shown for the $18$ redshifts planes listed in Table  \ref{table:redshifts}.
Six `groups' of lines can be clearly identified on Figure \ref{fig:l2Cl_fracbox_angles}. Each group corresponds to one of the six box sizes. Within each group, the 18 redshifts give very similar results, it is difficult to distinguish them.
As expected, the theoretical calculations predict that missing $k$-modes do not affect measurements for 
$\ell / \ell_{box} > 1$.
For $\ell / \ell_{box} < 1$, the measurements from the different redshifts source planes can be superimposed with a scatter of no more than a few percent. The results for HF1 and HF2 are very similar and not shown here.

Although it was known that periodic replica of the simulation boxes in N-body simulations lead to power suppression at large angles, and here we have a quantitative measurement of this loss, which, to the best of our knowledge, has never been done before.

\begin{figure*}
   \centering
    \includegraphics[width=2.5in]{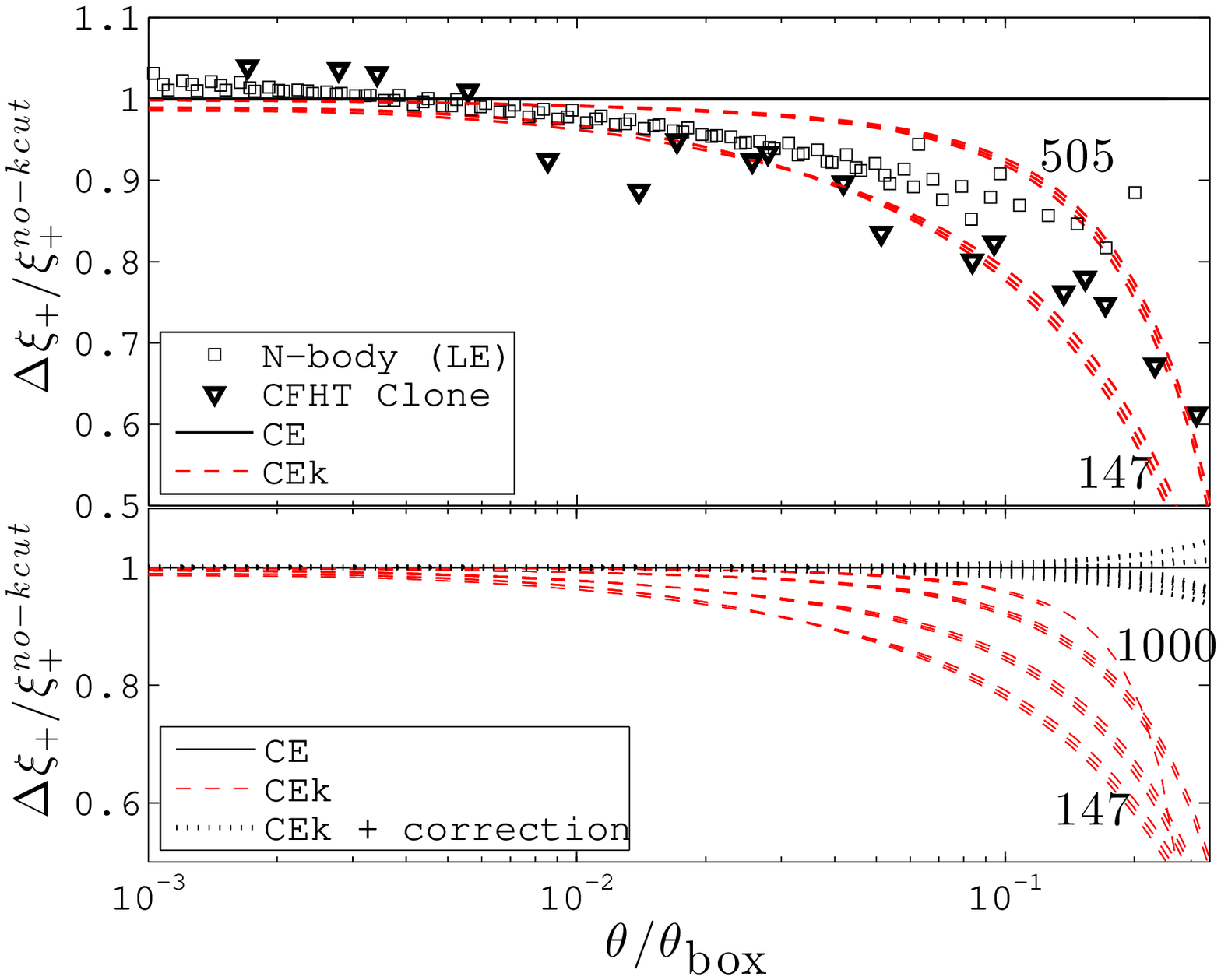} 
    \includegraphics[width=2.5in]{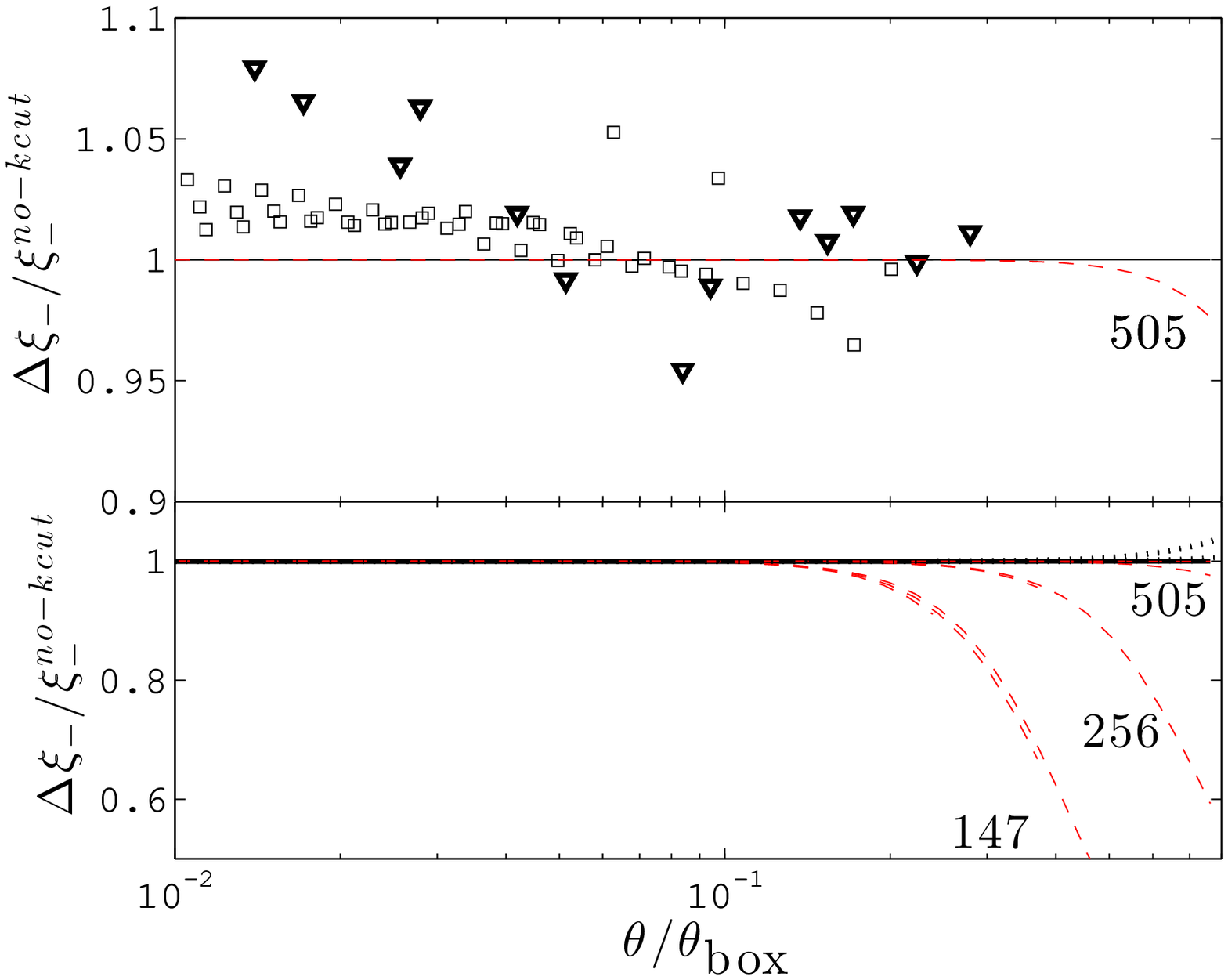} 
   \caption{(Top left:) Ratio between the measurements of $\xi_{+}$ in the CE and CEk models. 
   As for Fig. \ref{fig:l2Cl_fracbox_angles}, we over-plot the measurements different  sources planes and $L_{var}$.
     The $x$-axis is the ratio between the measurement angle $\theta$ and $\theta_{box}(z)$ (i.e. the angle subtended by the simulation box at the source redshift, see main text).
   The lines labeled $147$ and $505$ are actually stacked results from $z_s = 0.582$, $0.968$ and $3.084$, which align very well.   
  Also shown are stacked measurements from the LE suite (with $z_s = 0.582$, $0.968$ and $3.084$, $L_{box} = 505\mathMpch$) and from the CFHT clone (with $z_s = 0.5$ and $1.0$, $L_{box} = 147 \mathMpch$). 
   (Bottom left:) Same models as the top panel, but with the addition of $L_{var} = 256$ and $1000\mathMpch$.
We also present the `corrected' measurements, seen in the top right region of the panel, which use equation \ref{eq:f_ell} to undo the effect of the low-$k$ cuts
prior to the integration of equation \ref{eq:xi}.           
   (Right:) Same as left panel, but for $\xi_{-}$.}
      \label{fig:xip_fracbox}
\end{figure*}

The redshift dependence of the  {suppression} is mostly {dependent on} the ratio $\ell/\ell_{box}$, which suggests that for a fixed $L_{box}$, the  {FS} effect at all angles and all redshifts (equation \ref{eq:C_ell_lost}) can be captured with very few parameters.
We chose to describe it with a simple power law:
 \begin{eqnarray}
 \frac{C_{\ell}^{k-cut} }{ C_{\ell}} -1  = \begin{cases} 10.0 [ (\ell / \ell_{box})^{a(L_{box})} - 0.992]   &\mbox{if } \ell/\ell_{box} < 1.0 \\ 
                             0 & \mbox{otherwise} \end{cases} 
  \label{eq:f_ell}
 \end{eqnarray}
{Since the actual curves become shallower for  $\ell/\ell_{box} \ll 1.0$, 
we instaure a limit to avoid over-correcting the effect. We find that a correction ceiling at 70 percent  offers a good trade off
between under- and over-correction (this choice will be justified quantitatively at the end of next Section)}.
The factor of $0.992$ vertically shifts the $y$-intercept of the power law, which serves as a simple way to take into account 
 a smooth transition that occurs  in the range $0.7 < \ell / \ell_{box} < 1.0$.
This {fit} is shown by the red dashed lines on Figure \ref{fig:l2Cl_fracbox_angles}, its accuracy is better than 5 percent.

 The final power spectrum correction is therefore given by:
 \begin{eqnarray}
 f^{corr}_{\ell} \equiv  max(min(\frac{C_{\ell}^{k-cut} }{ C_{\ell}} -1 ,0) , -0.7) 
 \label{eq:f_ell_corr}
 \end{eqnarray} 
 For the smallest box, the fit describes accurately the drop only for $\ell/\ell_{box} > 0.4$, while it is valid down to $0.2$ for the largest three boxes. 
The dependence of $a$ on $L_{box}$ is shown in Fig. \ref{fig:a_lbox}, which exhibits a clear power law scaling with
$a(L_{box}) =  0.58 L_{box}^{-0.403}$. 

With this at hand, we can now compute accurately the exact value of $C_{\ell}^{lost}$
and calibrate the measurement from any simulation at any redshift.
 {The correction procedure can be summarized as follow:
 for a given $L_{box}$ and source redshift $z_s$, simply calculate $a(L_{box})$, $\ell_{box}(z_s)$ and $f^{corr}$.
 Then, with the equation given in this Section, any measured $C_{\ell}^{k-cut}$ can be `corrected' into a $C_{\ell}$ with no $k$ cut.}


\subsection{ $\theta$-space}
\label{subsec:trends2}

A similar analysis can be performed in real space for $\xi_{\pm}$ as a function of $\theta/\theta_{box}(z)$.
The top panels of Fig. \ref{fig:xip_fracbox} shows the amplitude loss of $\xi_{\pm}$ for the CE models and three different redshift planes. All measurements are normalized to CE.
The (HF1k/HF1) and (HF2k/HF2) ratios produce the exact same results and they are not shown.

As expected, the shear correlation function falls faster as the separation angle approaches the box size.
 With a box size of $L_{box}= 505\mathMpch$, $\xi_+(\theta)$ drops by (1, 10, 25 and 50) per cent for $\theta$ being
($1/100$, $1/10$, $1/5$ and  $1/4$) of the angular box size {$\theta_{box}(z_s)$}. The exact suppression factor of the shear correlation function is a strong function of $L_{box}$, as already shown in Fig. \ref{fig:l2Cl_fracbox_angles} in $\ell$-space.
The different slopes at low $\ell$, corresponding to different $L_{box}$,  generate different values of $\xi_+$ at large $\theta$.
The exact value depends on $a(L_{box})$, and the larger the  box size the smaller the effect.
At a tenth of a box, the $1 \mbox{Gpc}h^{-1}$ box misses about 5 per cent of signal, whereas the $147 \mathMpch$ box suffers from a  20 per cent loss.

Exactly as observed in $\ell$-space, the redshift dependence of the FS effect is hidden in the ratio $\theta/\theta_{box}(z)$.  
As shown in Fig. { \ref{fig:xip_fracbox}}, this trend is also observed in the simulations, where the  {3} different source redshifts stack on top of one another remarkably well. 
In addition, we over-plot the $z = 0.5$ and $1.0$ measurements from the TCS series \citep{2012MNRAS.426.1262H} 
that was used in the construction of the CFHTLenS mock catalogue \citep{2012MNRAS.427..146H}.
The TCS suite were constructed with a box size of $147\mathMpch$.
Once adjusted for the different cosmology of these simulations ({\it WMAP}5 + BAO + SN), the TCS signal follows the same  trend as  {CEk}.
This comparison supports the fact that the volume effect can be generalized to different cosmological models, with the redshift dependence hidden in $\theta/\theta_{box}(z)$ (or $\ell/\ell_{box}(z)$ in $\ell$-space).
The top-right panel of Fig. \ref{fig:xip_fracbox} shows the box size effect on $\xi_{-}$. It follows a similar scaling as for $\xi_+$, however the suppression occurs at larger separation angle.

The bottom two panels of Fig. \ref{fig:xip_fracbox} show how the signal for $\xi_\pm$ can be recovered using the correction proposed by equation \ref{eq:f_ell_corr}. This is shown by the thin dotted lines that scatter around the horizontal solid line that represents a perfect correction. The correction is accurate to better than 5 percent until roughly a third of the box size for $\xi_+$ and two thirds of the box size for $\xi_-$.  
  On this figure, all these dotted lines are in fact regrouping corrections to the six box sizes considered in this paper, all using equation \ref{eq:f_ell_corr}. 
 The choice of a cut at 70 per cent as the correction ceiling in equation \ref{eq:f_ell_corr} was chosen such that the group of lines would lie equally on both sides of the horizontal line (i.e.  a perfect correction).

\subsection{Summary of the FS correction for $C_{\ell}^{\kappa}$}
\label{subsec:FS_kappa}

The FS correction to the covariance of weak lensing observables extracted from general simulations with finite support
is summarized here. Ultimately, the goal is to obtain a reliable estimate of the two-point statistics covariance matrices $\mbox{Cov}^{\kappa}(\ell,\ell')$ and $\mbox{Cov}^{\xi\pm}(\theta,\theta')$ that can be used for precision cosmology using weak gravitational lensing data.

In Fourier space, the procedure is as follow:
\begin{enumerate}
\item{Compute the loss of power in the $C_{\ell}$ measurement due to the finite simulation box size, from equation \ref{eq:C_ell_lost}.}
\item{Correct for the effect in the covariance measurement: }
\end{enumerate}
\begin{eqnarray}
\mbox{Cov}^{\kappa, FS}_{N-body}(\ell,\ell') = \mbox{Cov}^{\kappa}_{N-body} \times (1 + [1/C_{\ell}^{lost}])^2 \delta_{\ell\ell'} \nonumber \\
\equiv \mbox{Cov}^{\kappa}_{N-body} \times FS^{\kappa}.  
\label{eq:FS_kappa}
\end{eqnarray}
This mostly corrects the largest scales, hence a Gaussian estimate is being assumed for the error on $C_{\ell}$.
Accordingly, the Kronecker function ensures that the correction is applied exclusively on the diagonal component. 
Alternatively, one can use the power law approximation for $f^{corr}$,  combined with equation \ref{eq:f_ell_corr},
 to construct $C_{\ell}$ from the $C_{\ell}^{k-cut}$ that are measured from the $N$-body suite. 

The following Section is dedicated to the recalibration of the covariance matrix $\mbox{Cov}^{\xi +}(\theta,\theta') $ {at small scales,} taking into account non-Gaussian effects.

\section{Recalibration of $\mbox{Cov}^{\xi \pm}(\theta,\theta')$}
\label{sec:cov}

In the last Section we have quantified how the super modes affect the two-point functions in real and Fourier space,
and proposed a simple parameterization to correct for the amplitude suppression on $C^{\kappa}_{\ell}$ and $\xi_{\pm}(\theta)$.
In this Section, we focus on the large and small scale recalibration of the shear correlation function covariance matrices $\mbox{Cov}^{\xi \pm}(\theta,\theta')$ 
{. These two opposite regimes are affected} by the finite box effect and the non-linear clustering respectively. 
A specific recalibration of the SLICS-LE simulation suite {is} proposed{, with a generalized prescription for any  
simulation suite}.
Issues related to {\it beat coupling} and {\it halo sampling variance} are ignored for the moment and will be discussed 
in  Section \ref{sec:application}.

\subsection{Large angle calibration of $\mbox{Cov}^{\xi +}(\theta,\theta') $}
\label{subsec:cov_xip}

\begin{figure}
   \centering
    \includegraphics[width=2.9in]{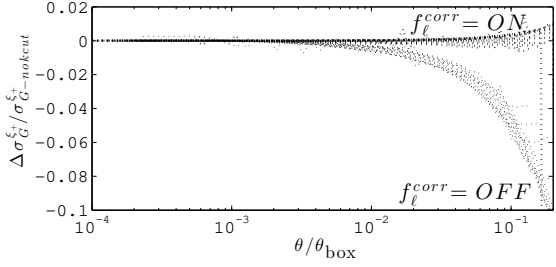} 
   \caption{Fractional error between the error estimates on $\xi_+$  with the HF2 and HF2k models.
   We plot these results against $\theta/\theta_{box}(z)$, and stack measurements for 15 different values of $z$ picked from Table \ref{table:redshifts}.
   These all superimpose as the group of lines labeled `$f_{\ell}^{corr} = OFF$'.
   We show the effect of the correction based on equation \ref{eq:f_ell} with the lines labeled  `$f_{\ell}^{corr} = ON$'.
   All the curves have been normalized 
   such that they asymptote to zero at small angles. We checked that choosing HF1/HF1k or CE/CEk  provide identical results.}
   \label{fig:Gauss_cov_xi_box}
\end{figure}

The elements in the real space shear covariance matrix are correlated, even at large separation angle where the density fluctuations are still linear and obey Gaussian statistics. The correction of the  {FS} effect on the covariance matrix at large scale must take into account the coupling of long wavelength modes. To this aim, we use the predictions and notation of \citet{2002A&A...396....1S}  (equations 32-34)  
to compute the Gaussian contribution to the covariance matrices on $\xi_+$.
In addition, we use an alternative estimator of  $\mbox{Cov}^{\xi\pm}(\theta,\theta')$
which is calculated directly from the multipole covariance matrix, as first derived by \citet{2008A&A...477...43J} in the Gaussian case, and later extended to the non-Gaussian case by \citet{2009MNRAS.395.2065T}\footnote{In these two references, the notation for the weak lensing power spectrum measured in $E/B$-modes is $P_{E/B}(\ell)$, but we label these quantities by  $C_{\ell}^{E/B}$  here for consistency
with the rest of the paper. }:
\begin{eqnarray}
   \langle \mbox{Cov}^{\xi\pm}(\theta,\theta') \rangle \propto \!\int\!\!\!\!\int\!\! \langle \mbox{Cov}^{\kappa}(\ell,\ell') \rangle J_{0/4}(\ell \theta) J_{0/4}(\ell'\theta')     \ell d\ell  \ell' d\ell'  
   \label{eq:Cov_xipm_joachimi}
\end{eqnarray}
The equivalence between both methods is established in the large angle limit,
but generally differ at smaller angles. We follow the empirical prescription of \citet{2011ApJ...734...76S}
and rescale equation \ref{eq:Cov_xipm_joachimi}  by a multiplicative pre-factor 
of the form $1/f(A,z)$ to account for the discrepancies.
This can be written as $f(A,z) = max(\alpha_z/A^{\beta_z},1.0)$,
with $\alpha_z = 3.2952 z^{-0.316369}$ and
$\beta_z = 0.170708 z^{-0.349913}$.
This correction factor was calibrated for a slightly different cosmology, but this would have only a marginal effect on our calculations.
 
For the 15 highest redshifts of Table \ref{table:redshifts}, Fig. \ref{fig:Gauss_cov_xi_box} shows the relative error of the $\xi_+$ diagonal error bars with and without the large scale $k$ cuts, i.e. $\Delta \sigma^{\xi+}_G / \sigma^{\xi+}_{G-nokcuts}$, where $ \sigma_{G}^{\xi+} = \sqrt(\mbox{Cov}_{G}^{\xi+}(\theta =\theta'))$. 
The diagonal error bars were calculated following \citet{2002A&A...396....1S}.
{We observe that} the redshift dependence of the FS effect is again only a function of $\theta / \theta_{box}$; the scatter between the different redshifts is very small.
These calculations were computed assuming  $L_{box} = 505 \mathMpch$ and with the HF2 and HF2k models, but {HF1 and HF1k models} yield equivalent results. 

The ratio $\sigma^{\xi+}_G / \sigma^{\xi+}_{G-nokcuts}$ does not converge to unity at small scales  because the errors depend on the integral over $\xi+$.
 {The complete} effect of the finite box size is therefore  to produce a drop in signal at large angles,
 {plus an overall suppression} of about 2 per cent at all angles and redshifts. 
In Fig. \ref{fig:Gauss_cov_xi_box}, $\Delta \sigma^{\xi+}_G / \sigma^{\xi+}_{G-nokcuts}$ {is manually set to zero} at small $\theta / \theta_{box}$
 in order to  isolate the {dominating}  large angle drop. {As expected from Gaussian statistics}, we note that  the FS effect is the same for both
the signal and the error bar: for $L_{box} = 505 \mathMpch$, it causes a 10 per cent drop at $\theta / \theta_{box} = 0.1$.

The thin dashed lines scattered around the horizontal line shows the impact of applying the correction factor 
$f^{corr}_{\ell}$ {on the HF2k model}.
We observe that the error bars agree with the HF2 to better than a percent at all redshifts,
showing that the simple parameterization proposed in equation \ref{eq:f_ell} corrects accurately the box effects on the error.

\begin{figure}
   \centering
    \includegraphics[width=3.2in]{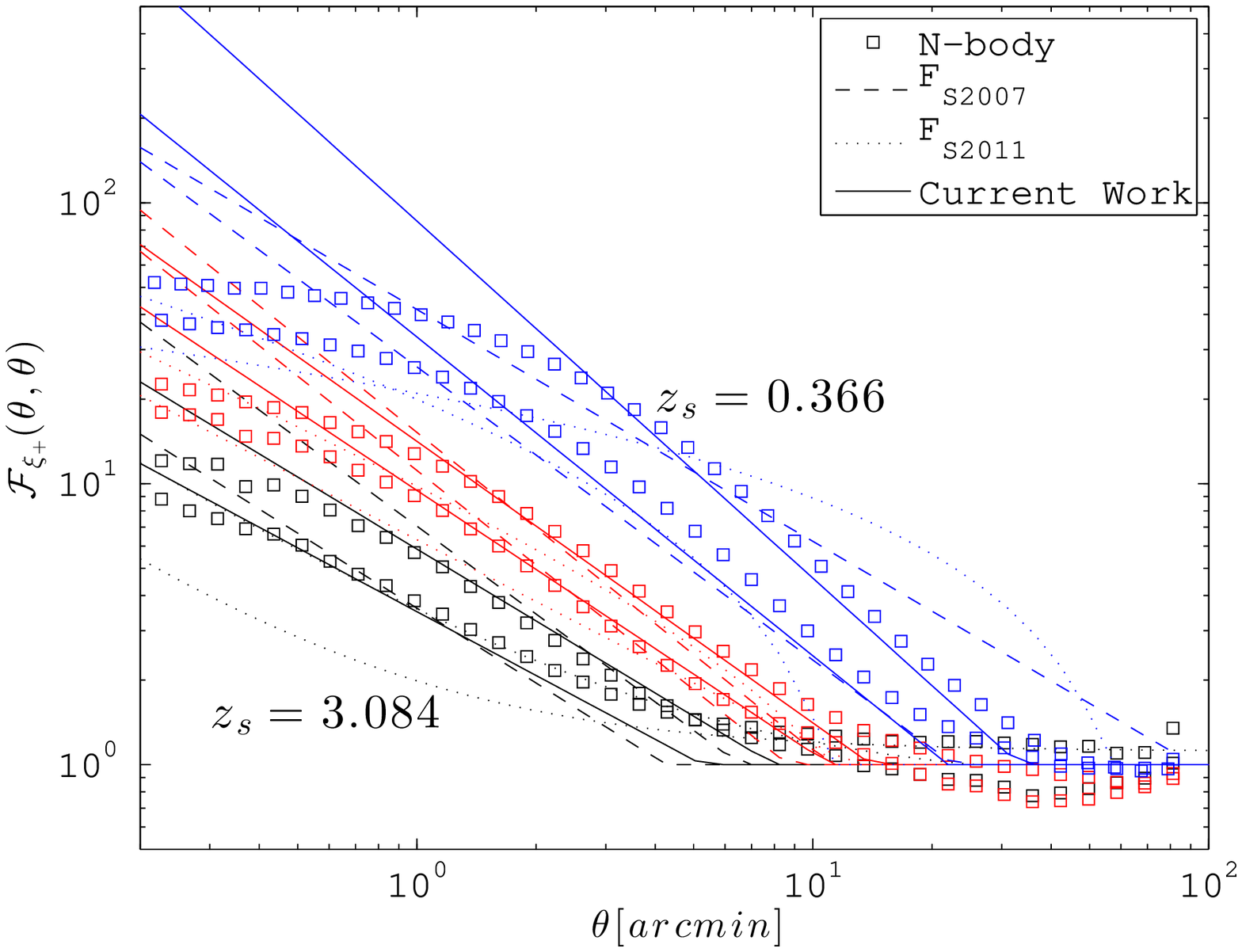}
    \includegraphics[width=3.2in]{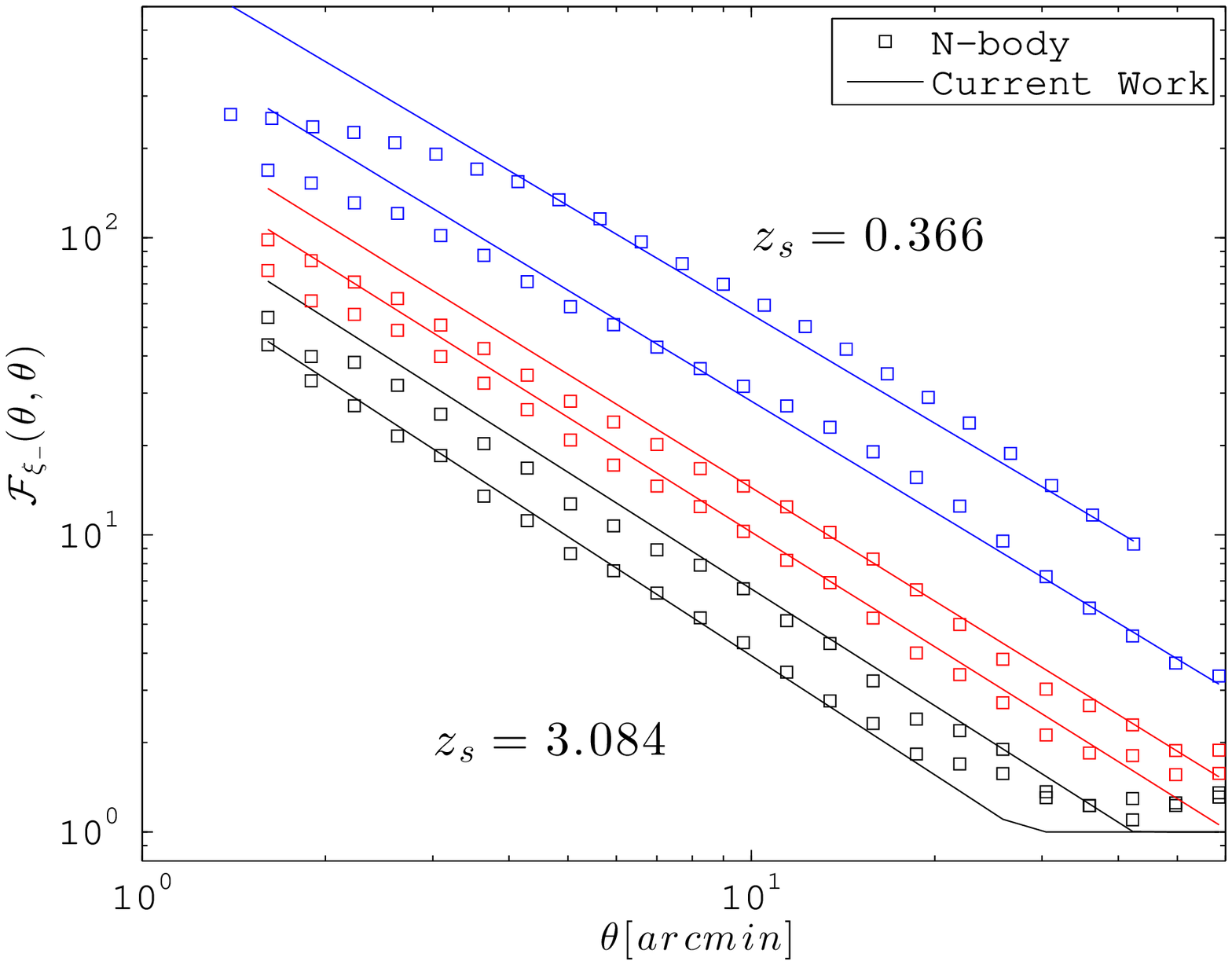}
     \caption{Diagonal elements of ${\cal F}_{\xi+}$ (top) and ${\cal F}_{\xi-}$ (bottom), as defined in equations \ref{eq:F_theta_new} and \ref{eq:F_theta_xim},  for  $z_s = 0.366$, $0.582$, $0.968$, $1.283$, $1.886$ and $3.084$. Lower (black) lines correspond to higher redshifts, higher (blue) lines are for lower redshifts. 
     The predictions from  ${\cal F}_{S2007}$, 
     ${\cal F}_{S2011}$ and from this work are shown in dashed, dotted and solid lines respectively.}
   \label{fig:F_theta_diag}
\end{figure}

 \subsection{Summary of the FS correction for $\xi_{\pm}$}
\label{subsec:FS_xi}

In real space, the FS effect can be corrected as follow:
\begin{enumerate}
\item{Compute the predictions for $\xi\pm$ from equation \ref{eq:xi},  
with and without the cut-off corresponding to the simulation box size in the lower bound of the $k-$integration. 
Note that this $k$ cut estimate can also be approximated from equation \ref{eq:f_ell_corr}, combined with the predictions for $C_{\ell}^{\kappa}$ without the $k$ cut,
followed by an integration over $\ell$ as in the left part of equation \ref{eq:xi}.}
\item{Compute the Gaussian predictions for $\mbox{Cov}^{\xi\pm}$ from the technique of \citet{2002A&A...396....1S},
both with and without the low-$k$ cut.}
\item{Correct the FS effect with:}
\end{enumerate}
\begin{eqnarray}
 \mbox{Cov}^{\xi\pm,FS}_{N-body} = \mbox{Cov}^{\xi\pm}_{N-body} \times \bigg[ \frac{\mbox{Cov}^{\xi\pm}_{G, no k-cut}}{\mbox{Cov}^{\xi\pm}_{G, k-cut}}\bigg] 
    \equiv \mbox{Cov}^{\xi\pm}_{N-body} \times FS^{\xi\pm}.
    \label{eq:FS_xi}
\end{eqnarray}

Generally, we can write the corrected covariance estimator $X$ as $\mbox{Cov}^{X,FS} = \mbox{Cov}_{N-body}^{X} \times FS^{X}$,
where the exact shape of $FS^{X}$ changes depending on the estimator ($\kappa$, $\xi_+$, $\xi_-$, etc.), 
but can generally be linked to $FS^{\kappa}$ in some way via the relation between most weak lensing estimators and the power spectrum.

{We emphasize once more that this correction mainly affects the largest scales of the simulated survey.}

\subsection{Small angle calibration of $\mbox{Cov}^{\xi_+}(\theta, \theta')$}
\label{subsec:fitfunctionxip}

An accurate modelling of  the full $\xi_{\pm}$ covariance matrix is an essential ingredient for cosmological parameters measurements and forecasting. The calculations from \citet{2002A&A...396....1S} are not valid in the non-linear regime. As shown in \citet{2007MNRAS.375L...6S}, the true variance is one to two orders of magnitude higher than the Gaussian case. Using a suite of $N$-body simulations, \citet{2007MNRAS.375L...6S} 
estimated the full non-Gaussian error:
\begin{eqnarray}
\mbox{Cov}^{\xi_{\pm}}(\theta,\theta') = \langle \Delta \xi_{\pm}(\theta) \Delta \xi_{\pm} (\theta')	 \rangle
\end{eqnarray}
which was expected to reconnect to the Gaussian limit at large angles.
They further measured the quantity:
\begin{eqnarray}
{\cal F}_{\xi+} (\theta, \theta') \equiv  \frac{\mbox{Cov}^{\xi_+}_{N-body}(\theta, \theta')}{ \mbox{Cov}_{G,[k-cut]}^{\xi_+}(\theta, \theta')}  
\label{eq:F_theta}
\end{eqnarray}  
which captures the departure from Gaussian calculations, and proposed a simple parameterization of ${\cal F}$.
This  allows to estimate the non-Gaussian error in weak lensing studies from linear theory. 
These results have to be revisited for the following reasons:

\begin{itemize}

\item{} First, their work was based on a cosmology with a high $\sigma_8$ compared to recently measured values: 
they used $\Omega_m = 0.3$, $\Omega_{\Lambda} = 0.7$, $\sigma_8 = 1.0$, $h = 0.7$,
and for this only, a re-calibration of the fit is valuable.

\item{} Second, the non-Gaussian errors were chosen to be parameterized
with a common power law, i.e. ${\cal F} _{S2007} \propto (\theta \theta')^{-\beta(z)}$,  
which is an strong over-simplification of the departure. 
This particular choice of parameterization was motivated by the shape of ${\cal F}_{S2007}$ on the diagonal terms, 
which traditionally weights maximally in the forecasts based on Fisher calculations. 
Unfortunately, there is no quantification regarding the performance and accuracy of the fit for the off-diagonal components, 
even though it is natural to expect a smooth shape from the high level of correlation. 
What we observe in the off-diagonal elements is incompatible with this fit, 
which is why  a different parameterization is necessary.

\item{} Third, the construction of their light cones is based on the tiling technique of \citet{2000ApJ...537....1W},
which involves stitching simulation boxes of decreasing size as we approach the observer.
They ran 2 simulations with $L_{box} =  800 \mathMpch$, three with $L_{box} = 600 \mathMpch$,
 four with $L_{box} = 400 \mathMpch$ and seven with $L_{box} = 200 \mathMpch$.
 Random rotations, box ordering and origin shifting were used on  16 independent simulations 
 to generate $64$ light cones.
They acknowledged the limitations of their results coming  from the fact that their light cones 
were not  independent,  but the setup adopted in this paper with the SLICS-LE simulations is a net improvement:
each simulation contributes to a single light cone, as opposed to four.

\item{} Fourth, the importance of finite simulation box sizes was not fully understood,
and therefore, in their calculations, \citet{2007MNRAS.375L...6S} did not include the $k$ cut in the denominator of equation \ref{eq:F_theta}. 
Consequently, they found that the function ${\cal F}_{S2007}$
was crossing unity at large angles, i.e. that non-Gaussian error measured from simulations 
dropped significantly {\it below} the Gaussian predictions. This feature was indeed attributed to the effect of measurements on finite support, 
but it now becomes clear that this can be modelled accurately. 

\end{itemize}

Many improvements to this fitting formula were provided by a second recalibration by \citet{2011ApJ...734...76S},
which recognized the importance of the super-sample modes in the reconnection between 
Gaussian and non-Gaussian measurements. Their approach was to create Gaussian realizations directly in the same simulation box 
-- which is equivalent to including a large scale  $k$ cut -- and to find the ratio of the non-Gaussian signal with this measurement.
Although much more accurate that the ${\cal F} _{S2007}$ estimator, the ${\cal F} _{S2011}$ is calibrated against Gaussian predictions
by \halofit2011, which we now know suffer from significant  loss of power at scales of a few arc minute. 
This inaccuracy  propagates both on the error about $\xi+$ and on its cross-correlation coefficients. 
The cosmology adopted in the ${\cal F} _{S2011}$ is not significantly different 
from that of the current work, but their fit function is not tested  for $z_s<0.6$, which
is unfortunate since many current and coming surveys have their source counts maximal at $z_s = 0.5$.


In the following we propose a new calibration of ${\cal F}_{\xi+}$ which resolves the issues with the older approaches mentioned previously. In order to increase the redshift sampling, we compute the Gaussian and non-Gaussian covariance matrices for all 18 redshift slices provided in Table \ref{table:redshifts}.
The top panel of Fig. \ref{fig:F_theta_diag} shows the matrices diagonal components for a few of these redshifts, compared to the results from \citet{2007MNRAS.375L...6S} and \citet{2011ApJ...734...76S}.
Except for $z_s = 0.366$, the ${\cal F}_{S2007}$ (dashed lines) and the simulations (squares) agree within 30 percent down to the arc minute. Most of the gain in the current calibration comes from the off-diagonals. 
The agreement with ${\cal F}_{S2011}$ is not as good, with significant departures at all angles and redshifts.

 At scales smaller than a few arc minutes, the $N$-body measurements depart from the power law, 
 but this is exactly where the progressive degradation of the resolution in the SLICS-LE  suite was flagged.
 We therefore exclude those scales in the recalibration. 
 
In order to decide on the  parameterization to adopt for the off-diagonal terms,  we first visualize the full function ${\cal F}_{\xi+}$; this is shown on Fig. \ref{fig:F_theta_offdiag} for $z_s = 0.582$. This characteristic bell shape appears for all redshifts.  The circular symmetry around the point $\theta=\theta'=0.1~{\rm arcmin}$ suggest that a good parametrization would be centred at $(0.1, 0.1)$ using polar coordinates, i.e.  $(\theta, \theta')  \rightarrow (R_{\theta}, \phi)$,
where $\phi$ represents the polar angle subtended between the coordinate pair and the $\theta$ axis. In log scale we use $\mbox{log} R_{\theta} = \sqrt{\mbox{log}^2(\theta/0.1) + \mbox{log}^2(\theta'/0.1)}$. With this new coordinate system,  we have $R_{\theta} = (\theta/0.1)^{\sqrt{2}}$, and the new parameterization for ${\cal F}_{\xi+}$ is given by:
\begin{eqnarray}
{\cal F} (\theta, \theta')   = {\cal F} (R_{\theta}) = \frac{\alpha(z)}{R_{\theta}^{\beta(z)}}
\label{eq:F_theta_new}
\end{eqnarray}  

With this {form}, fixing the diagonal elements also fixes the rest of the matrix. Note that if this symmetry is preserved for a different cosmology, this is a practical and simple task to extend our result to any cosmological model using only the fit along the covariance diagonal. The two parametric functions, $\alpha(z)$ and $\beta(z)$, are shown in Fig. \ref{fig:FitAlphaBeta} as a function of redshift.
The former is fitted with an offset exponential function in log-log space, the latter by a  power law: 
\begin{eqnarray}
\mbox{ln}\alpha(z) = a_1 \mbox{exp}[a_2\mbox{ln}z] + a_3 \mbox{ \hspace{5mm} and \hspace{5mm} } \beta(z) = b_1  z^{b_2}
\label{eq:paraFtheta}
\end{eqnarray}
Our best fit values  for this set of parameters are $(a_1, a_2, a_3, b_1, b_2)$ $=$ $(3.66, -0.408, -1.06, 0.994, -0.244)$,
with a fractional error generally below 10 percent and no more than 20 per cent for both $\alpha$ and $\beta$. 
The revised fits for ${\cal F}_{\xi+}$ is shown as the solid lines in the upper panel of Fig. \ref{fig:F_theta_diag}. The agreement with the simulations is much better, particularly where the physical scales are fully resolved within the $N$-body code. The agreement is better than 20 percent for $1 < [\theta, \theta'] < 20$ arc minutes on most off-diagonal elements. At low redshift, the largest departures occurs at very small scales, where the 
simulations are lacking resolution. Overall, the accuracy of our fitted covariance matrix is better than 30 percent for every matrix element down to 1 arc minute.

\begin{figure}
   \centering
    \includegraphics[width=3.2in]{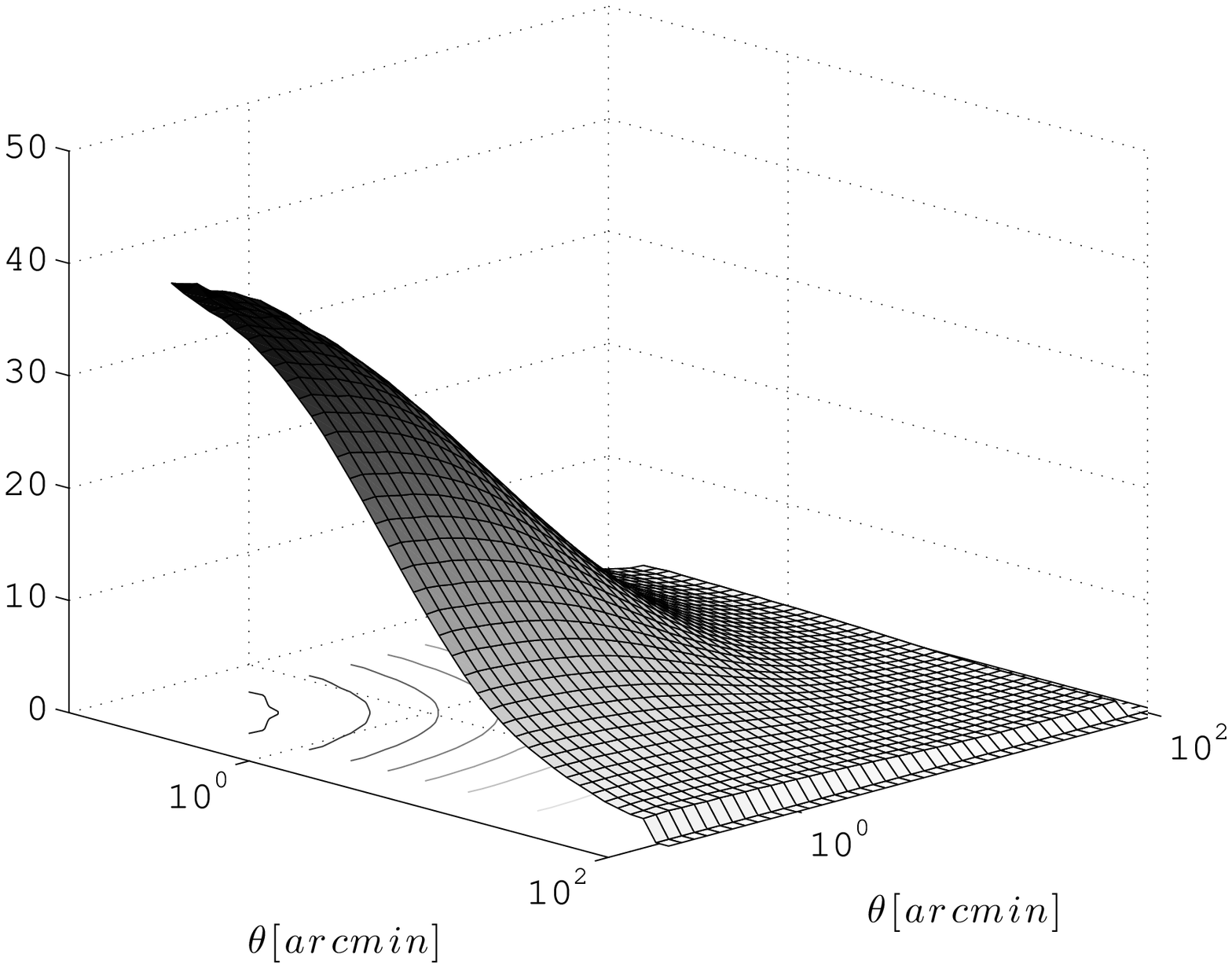}
    \includegraphics[width=3.2in]{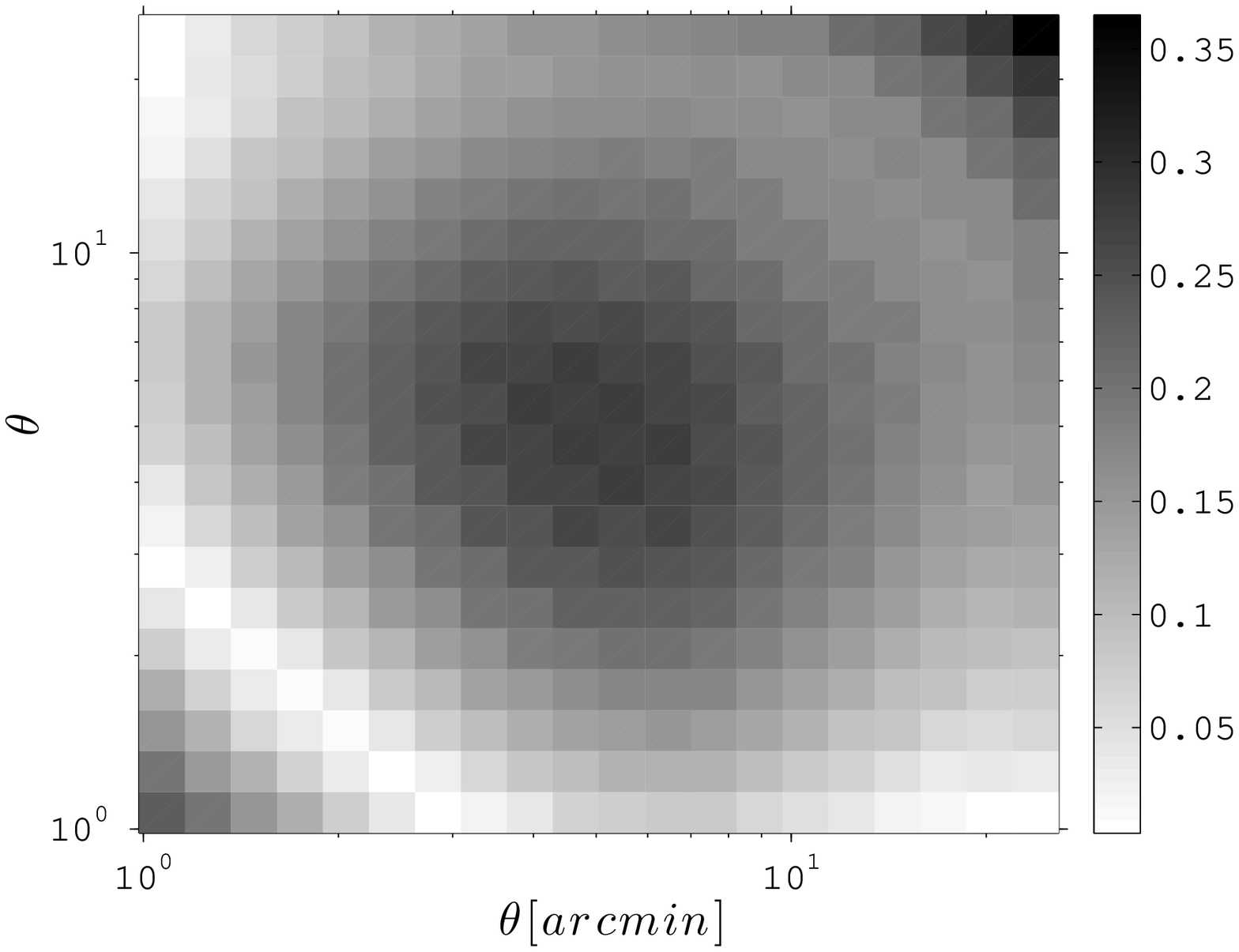}
     \caption{ (Top:) Full ${\cal F}_{\xi+}(\theta,\theta')$ measurement with $z_s = 0.582$.
     The  curves shown on the $x-y$ plane are projected lines of constant elevation. (Bottom:)
     Fractional error between ${\cal F}_{\xi+}(\theta,\theta')$ measured from the LE suite and constructed with the proposed fit.}
   \label{fig:F_theta_offdiag}
\end{figure}

\begin{figure}
   \centering
    \includegraphics[width=3.1in]{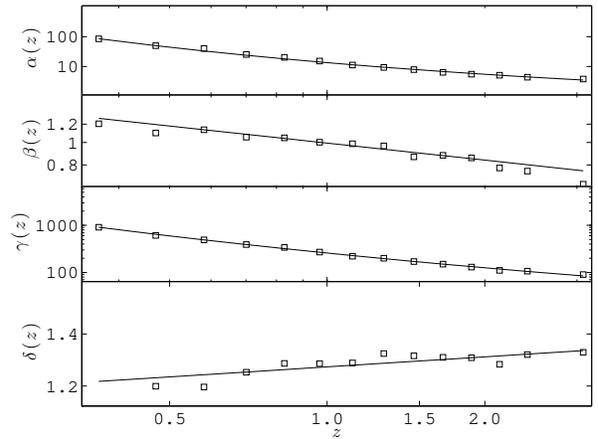}
     \caption{Measurements and fits to the parametric functions $\alpha(z)$,  $\beta(z)$, $\gamma(z)$ and $\delta(z)$ that enter in the modelling of ${\cal F}_{\xi\pm}$, 
     following equations \ref{eq:F_theta} and \ref{eq:F_theta_xim}.}
   \label{fig:FitAlphaBeta}
\end{figure}


\subsection{Small angle calibration of $\mbox{Cov}^{\xi_-}(\theta, \theta')$}
\label{subsec:fitfunctionsxim}

We carry in this Section the equivalent calibration for the $\xi_-$ quantity.
As mentioned in \citet{2011ApJ...734...76S}, the modelling of $\mbox{Cov}^{\xi_-}$
is significantly harder than for $\mbox{Cov}^{\xi_+}$ because of the complex
scale dependence of the Gaussian predictions, both on and off the diagonal.
These authors advocated that the non-Gaussian predictions should be 
calibrated directly against the $N$-body simulations instead,
but we argue here that a calibration matrix with good accuracy is still possible.

Fig. \ref{fig:FitXimCov} shows the full covariance matrix about $\xi_-$ measured from $N$-body simulations at $z_s = 3.084$,
compared to the corresponding Gaussian predictions. 
Results are obtained from equation \ref{eq:Cov_xipm_joachimi}, but the $f(A,z)$ correction was not applied since 
this empirical correction term is calibrated specifically for $\mbox{Cov}^{\xi_+}$.
We clearly see from the figure the reconnection between
both error estimates at large angles, which occurs close to the diagonal.
 Lower redshifts are overall similar, except that the reconnection between Gaussian and non-Gaussian estimates occur 
 at larger angles.    

An interesting result is that  no elements measured from the $N$-body suites appear to have negative values, in contrast with the Gaussian 
predictions \citep[see][]{2002A&A...396....1S}.
This is a strong hint that the anti-correlations present in the Gaussian case mostly disappear in the presence of non-linear mode coupling.

We agree with \citet{2011ApJ...734...76S} that a recalibration in the form of equation \ref{eq:F_theta}
is very challenging to model for the full matrix. 
However, based on  the smoothness and simplicity of the non-Gaussian surfaces, a simple modelling is still possible.

We define, on the diagonal only, 
\begin{eqnarray}
{\cal F}_{\xi-}  \equiv  \mbox{diag} \bigg[ \frac{\mbox{Cov}^{\xi_-}_{N-body} }{\mbox{Cov}_{G,k-cut}^{\xi_-}} \bigg]
\label{eq:F_theta_xim}
\end{eqnarray}  
The bottom panel of Fig. \ref{fig:F_theta_diag} shows ${\cal F}_{\xi-}$ for several redshifts.
Similarly to ${\cal F}_{\xi+}$, a single power law of the form ${\cal F}_{\xi-} (\theta) = \gamma(z)/\theta^{\delta(z)}$ seems to capture most of the non-Gaussian departure, 
down to the arc minute.
Fig. \ref{fig:FitAlphaBeta} shows the values of $\gamma(z)$ and $\delta(z)$ at $14$ redshifts using this fitting function. The functional form of $\gamma(z)$ and $\delta(z)$ with redshift $z$ is well described by:
\begin{eqnarray}
    \mbox{ln}\gamma(z) = g_1 \mbox{exp}[g_2\mbox{ln}z] + g_3 \mbox{ \hspace{5mm} and \hspace{5mm}}  \delta(z) = d_1 z^{d_2}
    \label{eq:paraFtheta_xim}
\end{eqnarray}
Our best fit values for this set of parameters are $(g_1, g_2, g_3, d_1, d_2)$ $=$ $(5.08, -0.221, 0.475, 1.27, 0.0429)$,
with an accuracy better than ten percent for all $z$. 
We show, in the bottom panel in Fig. \ref{fig:F_theta_diag},  that agreement between the 
 calibrated ${\cal F}_{\xi-}$ and the simulations measurement is strong.

From Fig. \ref{fig:FitXimCov}, we observe that the covariance matrix for $\xi_-$ at $z_s=3$
has a bell shape that peaks at $\theta_{peak} = \theta_{peak}' = 3$ arc minute with almost a circular symmetry around this peak. 
At lower redshifts, this peak occurs at larger angles ($\theta_{peak}=10$ arc minutes for $z_s=0.582$), but the same bell shape is observed.
 We find that we can model the full non-Gaussian surface by rotating the diagonal of $\mbox{Cov}^{\xi_-}$ about the peak, 
 using only values with $\theta \ge \theta_{peak}$.
The overall agreement between the measured and modelled $\mbox{Cov}^{\xi-}$ is better than 20 percent over most of the elements at all redshifts, however the far off-diagonal elements are generally too high by about 70 percent. 
Scales smaller than $\theta_{peak}$ are also affected by limitations in the resolution of the $N$-body simulations, hence we expect this region 
of the calibration matrix to be improved in the future, from simulations that resolve these scales more accurately.
Nevertheless, this method is simple to implement, and represents  six orders of magnitude improvement over the simple Gaussian calculations.

\begin{figure}
   \centering
    \includegraphics[width=3.4in]{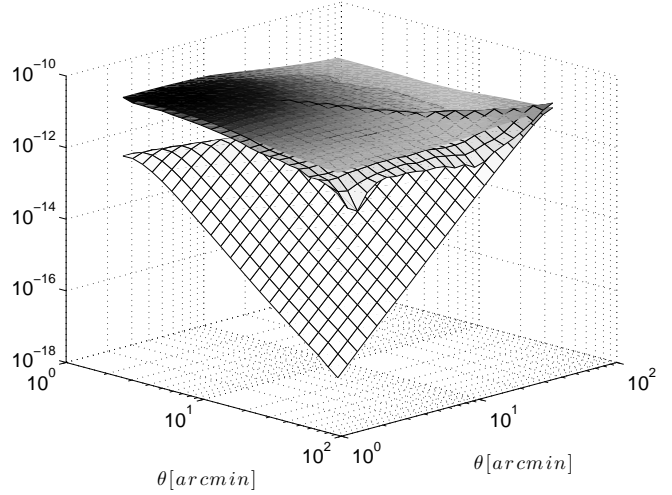}
\caption{ Covariance about $\xi_{-}$ at $z_s =  3.084$.
The top gridded surface represents the measurement from the $N$-body simulations, while bottom gridded surface shows the Gaussian prediction
computed with the HF2 model. Note that the $z$-axis is in logarithmic scale.
The third surface, plotted with no mesh pattern, is the result given by the parameterization of equation \ref{eq:paraFtheta_xim}. 
 Lower redshifts are qualitatively similar to this figure, although the connection  with the Gaussian prediction
occurs at larger angles and the non-Gaussian departure at small angles is amplified.}
   \label{fig:FitXimCov}
\end{figure}

\subsection{Summary of $\mbox{Cov}^{\xi\pm}(\theta,\theta')$ non-Gaussian calibration}
\label{subsec:summary_cov}

The prescription to construct the fully  non-Gaussian covariance matrix $\mbox{Cov}^{\xi+}(\theta,\theta')$, that properly correct for the FS effect, is given by:
\begin{enumerate}
\item{Compute the predictions for $\xi+$ from equation \ref{eq:xi},  
without applying the low-$k$ cut.}
\item{Compute the Gaussian error $\mbox{Cov}^{\xi+}_{G, no k-cut}(\theta,\theta')$ from the technique described in \citet{2002A&A...396....1S}.}
\item{Using equations \ref{eq:paraFtheta} and \ref{eq:F_theta_new} the diagonal components of the scaling matrix $\cal{F_{\xi+}}$ can be calculated.
The mapping between $\theta$ and $R_{\theta}$ is  $R_{\theta} = (\theta/0.1)^{\sqrt{2}}$, as described before  equation \ref{eq:F_theta_new}. 
If the sources are distributed in redshift, use the weighted mean of the distribution $\bar{z}$ (see Section \ref{subsec:nz}).}
\item{Construct the full ${\cal F}_{\xi+}$ matrix by spanning  rings in log-log space centred on $\theta = \theta' = 0.1'$, and assigning to these elements the value
found on the diagonal in step (iii).}
\item{Compute the full non-Gaussian matrix for $\xi_{+}$ from the relation $\mbox{Cov}^{\xi+}_{no k-cuts} = {\cal F_{\xi+}}\mbox{Cov}^{\xi+}_{G,no k-cuts}$.}
\end{enumerate}

The construction of $\mbox{Cov}^{\xi-}$ follows similar steps:
\begin{enumerate}
\item{Compute the Gaussian variance predictions for $\xi-$ from equation \ref{eq:Cov_xipm_joachimi} or from the method of \citet{2002A&A...396....1S}  (equations 32-34 therein).}
\item{From equation \ref{eq:paraFtheta_xim}, construct the diagonal components of the scaling matrix ${\cal F}_{\xi-} (\theta) = \gamma(z)/\theta^{\delta(z)}$. This can be
computed  either at the source  redshift, or, for a broad source distribution, using the weighted mean  redshift, as described in Section 6.1.}
\item{Construct the diagonal of the non-Gaussian matrix $\mbox{Cov}^{\xi-}$ from equation \ref{eq:F_theta_xim}.}
\item{Find the peak of this function and its corresponding angle ($\theta_{peak} = 3$'  at $z_s=3$ and 10' at $z_s = 0.58$) and rotate the diagonal elements about this maximum
to generate the matrix elements $\mbox{Cov}^{\xi-}(\theta' \ne \theta)$. For this step, only consider contribution from $\theta \ge \theta_{peak}$.}
\end{enumerate}

The cosmological dependence of ${\cal F}_{\xi\pm}$ is beyond the scope of this paper and will be  {investigated} in a future work.

\section{Application to Redshift Surveys}
\label{sec:application}

In this Section, we integrate the result of the previous section into a realistic framework that incorporates
extended distribution of sources and survey masks.

\subsection{Broad redshift distributions}
\label{subsec:nz}

Whether weak lensing is measured with a narrow or broad source redshift distributions, the true  distribution $n(z)$ is never a single redshift sheet. 
{It is therefore important to extend the previous work to a broad redshift distribution, extracted from the data and usually well described by analytic functions.}
 
This poses an additional complication in the modelling of the FS effect, since the total amount of power lost due to missing low-$k$ modes
is now a projection from multiple redshift slices, whereas our formulation depends on terms like $\theta_{box}(z_s)$ and $\ell_{box}(z_s)$.  
{These quantities are unfortunately not as clearly defined for a broad $n(z)$.}

{To test this,} we considered a distribution in the form of   
\begin{eqnarray}
\label{eq:nz}
   n(z) = \frac{\beta}{z_0 \Gamma(1+\alpha/\beta)} \bigg ( \frac{z}{z_0} \bigg)^{\alpha} \mbox{exp}\bigg[   -\bigg( \frac{z}{z_0}\bigg )^{\beta} \bigg].
\end{eqnarray}
We set the values of $\alpha$, $\beta$ and $z_0$ to (2.0, 1.58, 0.443) respectively, 
which provides a good match to the KiDS data \citep{2013Msngr.154...44J}.
This distribution peaks at about $z = 0.5$, and has a long tail at higher redshifts, with a median at $z = 0.8$ and weighted mean at $\bar{z} = \int  n(z) z  dz = 0.87$. 
We reproduce the calculations for each models of Table \ref{table:models}, and constructed 500 light cones from the SLICS-LE and -HR suites
with this $n(z)$ { inserted in equation \ref{eq:kappa_disc}}. 

We do recover a power suppression in both Fourier and real space, which fits well in our FS formulation if the source 
is taken to be at the weighted mean redshift $\bar{z}$.
The shape of the power loss  {stacks very well on top of the measurements at discrete $z_s$ shown in }  Fig. \ref{fig:l2Cl_fracbox_angles} and
 of Fig. \ref{fig:xip_fracbox}, meaning that we can use the  {prescription based on the } $f^{corr}_{\ell}$ correction term by employing $\bar{z}$
in  all the fitting function presented in this paper.
We also checked that the calibration matrices $\cal{F}_{\xi\pm}$ were consistent with the predictions at $\bar{z}$, both its shape and amplitude.
 {Generally, the effect of projecting different scales reduces the accuracy of our correction to the FS effect by no more than 5 percent, 
 compared to the discrete plane case.}

\subsection{Signal and error calibration with survey mask}
\label{subsec:correction}

Masking is unavoidable in gravitational lensing analysis because of the bright stars, satellite trails, etc., that need to be removed from the images.
We describe here how the address of FS effect can be adapted in the presence of a general survey mask.

\subsubsection{Mask applied on the mocks catalogues}

Masking is included in the mocks by applying the observed masks to the {simulated} light cones and  {weight} each mock galaxy (or pixel) 
 by the mask value. Once this is done, the mocks catalogues have both the $k$ cuts and the mask included, 
and as such should be compared to a theoretical model that also accounts for these two features.
The effect of masking is described as the product of the underlying density field and the mask in real space, 
or as a convolution of the Fourier space quantities. The convolution of the theory $C_{\ell}^{k-cut}$ with the masks power spectrum 
therefore gives the desired prediction, $C_{\ell, mask}^{k-cut}$,  and one can then use equation \ref{eq:xi} on this quantity to compute the model for $\xi_{\pm, mask}^{k-cut}$. 
This provides a consistent baseline between the masked mock catalogues and a theoretical signal, which is essential whenever 
one needs to test or calibrate a weak lensing estimator on the masked mocks.

Note that in this case, the FS effect is correctly accounted for, but no correction is applied. 
 Also note that the actual data is not affected by the $k$ cut, hence should be compared to $C_{\ell, mask}$ and $\xi_{\pm, mask}$.
 The calculation of two theoretical models -- with and without the $k$ cut -- is therefore necessary.
 In certain cases, it could be possible to {\it deconvolve} the mask from the mock observations, yielding an estimate 
 of $C_{\ell}^{k-cut}$ that can then be corrected with $f^{corr}_{\ell}$ and compared to the model  $C_{\ell}$ with no mask and no $k$ cut.


\subsubsection{Mask applied on the two-point function}

One of the advantage of the mock catalogue over real data is that the masking can be added or removed at will, allowing for a careful
understanding of its impact on the measurement. As an alternative to weighting the mock galaxies, one can also conduct the two point function
analysis on the simulations {\it without} the mask, then integrate the effect of masking on the two point function afterwards. 
This becomes advantageous notably for $C_{\ell}$ measurements, since, as an intermediate step,  
one can use the tools provided in this paper to undo the FS effect from the unmasked mocks  -- with the $f^{corr}_{\ell}$ correction term --
and convolve the corrected measurement with the mask subsequently. This would make the measurement from the simulations fully consistent with the data,
and both could be compared to a unique prediction, i.e. $C_{\ell, mask}$ (with no $k$ cut).
For the $\xi_{\pm}$ measurements, this is even simpler since the mask has no effect on the mean, only on the error.
It is therefore only a matter of correcting the mocks for the FS effect, then one can directly compare the data and the simulations with the $\xi_{\pm}$ model (with no $k$ cut).


\subsubsection{Mask and covariance}

The covariance matrix calculations in presence of a survey mask contain a higher level of complexity: 
in addition to  the FS effect described in this paper, two complimentary finite box contributions
must be included: the {\it halo sampling variance} (HSV) and the 
 {\it beat coupling} (BC). We briefly review the origin of these two quantities, then expose our strategy to incorporate them all
 in a consistent manner.
 
The HSV term comes from the finiteness of the simulation volume which
 imposes a constant density at the simulation box scale (this is the same across all realizations). A larger box size, $\widetilde{L}_{box} \gg L_{box},$ allows fluctuations at the scale of $L_{box}$, which in turn produces more massive halos and larger voids. Their absence results in a {\it missing} variance in simulation with smaller box size,
 a quantity labeled $\mbox{Cov}^{\kappa,HSV}$  that can be calculated analytically following \citet{2009ApJ...701..945S}, 
  then added to the sampling variance measured from the  simulations with $L_{box}$. 

The BC term, first identified in \citet{2006MNRAS.371.1205R} and \citet{2006MNRAS.371.1188H}, 
comes from the interaction between the survey mask and the `true' covariance.
In $N$-body simulations this interaction vanishes,  hence an error estimate based on unmasked mocks will underestimate the `true' error.
The correct  error, $\mbox{Cov}^{\kappa,obs}$, can be written as a two-dimensional convolution of the `true' covariance with the mask power  \citep{2009MNRAS.395.2065T}, 
and the BC contribution is generally represented by $\mbox{Cov}^{\kappa,BC} =   \mbox{Cov}^{\kappa,obs} - \mbox{Cov}^{\kappa,true}$.  
Its contribution on $P(k)$ at the survey box scale can be modelled analytically  to better than 10 per cent accuracy \citep{2014arXiv1401.0385L},
which can then be propagated  onto  $\mbox{Cov}^{\kappa, BC}$ as in \citet{2009MNRAS.395.2065T}.

The HSV and BC contributions can both be computed analytically and then included in the error budget 
as extra additive terms to the covariance  about $C_{\ell}^{\kappa}$.   
In the light of the current work, it is now clear that simply adding these contributions to the uncorrected simulation covariance gives an 
 incomplete account of the full box effect: the error extracted from mock catalogues needs to be rescaled to include the FS contribution. 
 To the best of our knowledge, all box effects can be incorporated in the  covariance with:
\begin{eqnarray}
  \mbox{Cov}^{\kappa,tot} = \mbox{Cov}^{\kappa, FS}_ {N-body} +\mbox{Cov}^{\kappa,HSV} + \mbox{Cov}^{\kappa,BC}
  \label{eq:total_error}
\end{eqnarray}
where
\begin{eqnarray}
\mbox{Cov}^{\kappa, FS}_{N-body} = \mbox{Cov}^{\kappa}_{N-body}\times FS^{\kappa}
\end{eqnarray}
Here,  the $FS^{\kappa}$ factor is the new {\it finite support} term central to this paper: it corrects for the effect of the low $k$ cut on  4-point functions
 (and the 2-point functions as well, see equation \ref{eq:FS_kappa}).

To estimate the error on real space quantities, one can then use equation \ref{eq:Cov_xipm_joachimi} 
on each term of equation \ref{eq:total_error} to produce $\mbox{Cov}^{\xi\pm, HSV}$ and $\mbox{Cov}^{\xi\pm, BC}$ as in  \citet{2009MNRAS.395.2065T},
but also $\mbox{Cov}^{\xi\pm, FS}_{N-body}$. This last term is a bit trickier: simulated light cones do not cover the full sky,
while this calculation involves an integral over {\it all} $\ell$.
One needs to include the contributions from $\ell<\ell_{lightcone}$ by grafting a Gaussian covariance
matrix  as done in \citet{2013MNRAS.430.2200K}. Although completely equivalent, it seems simpler  
 to compute $\mbox{Cov}^{\xi\pm, FS}_{N-body}$ as in equation \ref{eq:FS_xi}, i.e.  by correcting the simulation measurements directly in real space. 

Since the masks reduce the effective area of the survey, one must finally scale up the total error by the ratio between the unmasked area and the simulated light cones. 

One ingredient that is currently missing from the error calculation is the interaction between the mask and the covariance 
{\it at small scales}. The masking procedure  inevitably introduces an extra level of non-Gaussianity, and theses are not accounted  for in the current prescription.
 It is hard to predict the significance of this contribution, but 
 we intend to follow the approach of \citet{2012MNRAS.423.2288H} 
and quantify its importance in a  future work. 

\section{Conclusion}
\label{sec:conclusion}

Simulations are central to weak lensing data analyses for calibration and verification of estimators, for studies of systematics linked to secondary signals,   but also
  to provide an accurate description of the errors about the measurement.
In this paper, we investigate the impact of finite box size on weak lensing measurements performed in simulations
and  identify a new contribution to the uncertainty that has been overlooked in the past, which we coin the {\it finite support} (FS) effect.  
This contribution arises from the missing super sample modes that produce a suppression of the two point function at large scales, 
which leaks into higher order statistics.

We predict the impact of the FS effect for measurements of $C_{\ell}^{\kappa}$, $\xi_+$ and $\xi-$,
and propose simple recipes to rescale the measured signal and covariance.
The rescaling factors primarily depend on the simulation box size $L_{box}$ and on the ratio $\theta/\theta_{box}(z_s)$ (or $\ell / \ell_{box}(z_s)$), 
but are independent of the choice of theoretical model -- minimal variations are observed between \halofit2011, \halofit2012 and the Cosmic Emulator.
We verify these calculations against two series  of $N$-body simulations:
the new SLICS-LE suite with $L_{box} =505\mathMpch$, and the TCS suite  --  
used for the CFHT Clone -- which has $L_{box} =147\mathMpch$.

The  lensing power spectrum $C_{\ell}^{\kappa}$ is negligibly affected by the FS effect as long as the physical scales that are probed are fully contained within the simulation volume. For the largest scales and highest redshift, simulated light cones escape the volume, causing the amplitude of the power spectrum to drop. 
This effect can be accurately captured analytically by simple functions of  $L_{box}$
and is easily corrected in the signal (see equations \ref{eq:f_ell} and \ref{eq:f_ell_corr}) and in the covariance (see equation \ref{eq:FS_kappa}).

Real space  quantities  like $\xi_{\pm}(\theta)$, however, are more sensitive to box effects,
even at scales well inside the simulation box. This can be understood from their dependence
on the  {\it integral} over the power spectrum, causing any missing large scale $k$-modes
to affect a wide range of angles. 
Specifically, if the ratio $\theta/\theta_{box}(z_s)=0.1$, $\xi_{+}(\theta)$ and its error are suppressed by 5, 10, 20 and 25 percent for 
$L_{box}= 1000$,  $500$, $250$ and $147 \mathMpch$ respectively, independently of the source redshift. 
For $\theta/\theta_{box}(z_s)=0.2$, the suppression exceeds 25 percent even for $L_{box} = 1$ Gpc. 
With our simple parameterization (see equation \ref{eq:FS_xi}),  we can undo this FS effect with high fidelity, both in the signal and the error:
the residual differences between the corrected simulations and the continuous theory model (i.e. with no missing large scale modes) 
are generally of a few per cent only at all angles, even for volumes as small as those used for the CFHT Clone.

We discuss how all known finite box effects might be incorporated in a weak lensing data analysis,
in the presence of a broad distribution of source redshift and including survey masking.
In the light of these results, we find that box effects on covariance matrices can be fully modelled,  
and thereby advocate for an estimation strategy based on large ensembles of simulations with sub-Gpc volume
 that accounts and corrects for finite box effects.

Finally, we  propose a revised calibration matrix ${\cal F}_{\xi+}$ that maps Gaussian calculations onto non-Gaussian covariance estimates about $\xi_{+}$,
which are essential for accurate  forecasting and parameter extraction based on MCMC methods.
%
Our fit involves only 5 numbers that we determined with 15 redshifts checkpoints from the simulations,
and is confirmed to hold an element-by-element accuracy of 20-25 percent for $z<3$ and $\theta > 1.0'$,
even in the far off-diagonal regions.
We further present the first  parameterization of the non-Gaussian covariance estimates about $\xi_{-}$,
which also involves only 5 numbers and is 20 percent accurate over the diagonal and  most of the $(\theta,\theta')$ plane,
up to $z=3$ and $\theta > 1.0'$.

Many ideas  presented in this paper could easily be extended to other fields of  cosmology that rely on simulations, 
notably cross-correlation studies of weak lensing maps with galaxy fields or other tracers of matter. 
The FS corrections play a key role to guarantee the accuracy of calculations based on simulated mock catalogues.
So far, the  modelling of the FS effect has been tested on $C^{\kappa}_{\ell}$, $\xi_{\pm}$ and their respective covariances,
and extensions to other common cosmological observables (angular clustering functions, aperture mass, etc.) should be straightforward. 
It is our hope that such extensions will become available in preparation for future large surveys.

\section*{Acknowledgements}

The authors would like to thank Chris Blake for his comments on an earlier version of the manuscript,
and acknowledge significant discussions with Benjamin Joachimi and Masahiro Takada  about finite box effects in general.
Computations for the $N$-body simulations were performed on the GPC and TCS supercomputers at the SciNet HPC Consortium. 
SciNet is funded by: the Canada Foundation for Innovation under the auspices of Compute Canada; 
the Government of Ontario; Ontario Research Fund - Research Excellence; and the University of Toronto. 
JHD is supported by a CITA National Fellowship and NSERC, and LvW is funded by the NSERC and Canadian Institute for Advanced Research CIfAR.
This work was supported in part by the National Science Foundation under Grant No. PHYS-1066293 
and the hospitality of the Aspen Center for Physics.

\bibliographystyle{hapj}
\bibliography{mybib3_new}

\bsp

\label{lastpage}

\end{document}

%% file: mydefs.tex
\def\Mpch{$h^{-1} \mbox{Mpc} $}      
           
 \def\mathMpch{h^{-1} \mbox{Mpc} }      
 \def\mathhMpc{h \mbox{Mpc}^{-1}}

 \def\mathGpch{h^{-1} \mbox{Gpc} }

\def\halofit{{\small HALOFIT}}          
\def\cubep3m{{\small CUBEP$^3$M}}